\documentclass[aps,prb,twocolumn,groupedaddress, amsmath,amssymb,showpacs,
eqsecnum,floatfix]{revtex4}

\usepackage{graphicx}
\begin{document}


\title{Phase diagram of the vortex system in layered superconductors
with strong columnar pinning}

\author{Chandan Dasgupta}
\email[]{cdgupta@physics.iisc.ernet.in}
\altaffiliation{Also at Condensed Matter Theory Unit, Jawaharlal Nehru Centre
for Advanced Scientific Research, Bangalore 560064, India.}
\affiliation{Centre for Condensed Matter Theory, 
Department of Physics, Indian Institute of Science, 
Bangalore 560012, India}
\author {Oriol T. Valls}
\email[]{otvalls@umn.edu}
\affiliation{School of Physics and Astronomy and Minnesota Supercomputer
Institute, University of Minnesota, Minneapolis, Minnesota 55455}

\date{\today}

\begin{abstract}
We present the results of a detailed investigation 
of the low-temperature properties of the vortex system in strongly
anisotropic layered superconductors with a random array 
of columnar pinning centers.
Our method involves numerical minimization of a free energy
functional in terms of the time-averaged local vortex density. It yields
the detailed vortex density distribution for all local free-energy
minima, and therefore allows the computation of any desired  correlation
function of the time-averaged local vortex density. 
Results for the phase diagram in the temperature vs. pin concentration 
plane at
constant magnetic induction are presented. We confirm that for very
low pin concentrations, the low-temperature phase is a Bragg glass, 
which melts into an interstitial liquid phase via two first-order
steps, separated by a Bose glass phase. At higher concentrations, however,
the low-temperature  phase is a Bose glass,
and the melting transition becomes continuous. The transition
is then characterized by the onset of percolation
of liquid-like regions across the sample. Inhomogeneous local melting
of the Bose glass is found to occur. There is also a 
depinning crossover between the interstitial liquid and a completely
unpinned liquid at higher temperatures. At sufficiently large pin
concentrations, the depinning line merges with the Bose glass to 
interstitial liquid  transition. 
Many of the features we find have been observed experimentally 
and in simulations. We discuss the implications of our results
for future experimental and theoretical work.
\end{abstract}


\maketitle

\section{Introduction\label{intro}}

Understanding the
effects of random pinning on the properties of the mixed phase 
of high-temperature superconductors (HTSC) and other type-II superconductors 
is of fundamental importance in the
development of a theory of the effects of quenched disorder on the
equilibrium and dynamical behavior of a collection of interacting objects
(vortex lines or pancake vortices in the case of superconductors in the
mixed phase). This subject is also of great practical importance
because most  applications of HTSCs require a large critical
current, which can be achieved only in the presence of effective pinning.

Columnar pinning arising from damage tracks produced by
heavy-ion bombardment has received much attention in this context
because such extended defects parallel to the direction of the average
magnetic flux are very effective~\cite{Civale91,Budhani92}
in increasing the
critical current by localizing vortex lines along their length.
Heavy-ion irradiation produces  parallel
columnar defects, in  a random
configuration, each of which can trap one or more vortex lines at low
temperatures. The effects of an array of such extended defects,
oriented perpendicular to the superconducting layers, on
the properties of the mixed phase of HTSCs in a magnetic field
perpendicular to the layers have been extensively
studied  experimentally
~\cite{Civale91,Budhani92,Budhani94,Jiang94,Khaykovich98,Banerjee03,Menghini03,Banerjee04}, 
theoretically~\cite{NV93,Radz95,LV95,LV04}, and 
numerically~\cite{Sugano98,Sen01,TG03,NH04}. In
the regime in which the areal concentration of 
randomly placed columnar pins exceeds that
of vortex lines (as determined by the value of the magnetic
induction $B$) the behavior
of this system is now fairly well-understood. Theoretical 
studies~\cite{NV93,LV95} based on a mapping of the
thermodynamics of the system to the quantum mechanical properties of a
two-dimensional system of interacting bosons in a random potential, as well
as experimental~\cite{Budhani94,Jiang94}
studies, have established that in this regime the system undergoes
a continuous phase transition between a low-temperature 
``strong'' Bose glass (BoG) phase
and a high-temperature vortex liquid phase. In the
``strong'' BoG phase nearly all the vortex lines are strongly
pinned at columnar defects and both translational and bond-orientational
correlations of the vortex lines are short-range, indicating a
strongly disordered structure.

The behavior of the system in the ``dilute pin'' regime, where 
the number of columnar
pins is much smaller than the number of vortex lines, is much less 
well understood
.
Several years ago, it was pointed out by 
Radzihovsky~\cite{Radz95} that in this regime the system may exhibit a 
first-order transition between a low-temperature ``weak'' BoG and a
high-temperature vortex liquid. In the ``weak'' BoG phase only a 
fraction of vortex lines are strongly pinned at the columnar defects, and the
remaining vortex lines are weakly pinned in the interstitial region between 
defects. The liquid phase into which the ``weak'' BoG melts upon
increasing the temperature is expected to have a mixed character in the
sense that some of the vortices remain pinned at the pinning centers,
while the other, interstitial ones form a liquid. This partially pinned  
interstitial liquid (IL) should then exhibit a crossover to a 
completely depinned vortex
liquid at a higher temperature. This behavior is also suggested in a recent
theoretical study~\cite{LV04} based on the analogy of the vortex
system with a two-dimensional system of bosons in a random potential.

Recent experiments~\cite{Banerjee03,Menghini03,Banerjee04} on the highly
anisotropic, layered HTSC ${\rm Bi_2Sr_2CaCu_2O_{8+x}}$ (BSCCO) with a small
concentration of randomly placed columnar pins perpendicular to the layers
have provided many interesting results, some of which are in agreement
with the theoretical predictions mentioned above. These experiments 
were carried out for different values of the relative pin 
concentration $c$, ($ c= N_p/N_v \equiv B_\phi/B$, where 
$N_p$ is the number of columnar pins, $N_v$ the number of vortex lines
and $B_\phi$ the 
``matching field''). For
small values of $c$, the experiments indicate a first-order phase 
transition between a high-temperature vortex liquid and a
low-temperature BoG phase in which the vortices are arranged in a 
polycrystalline structure with grain boundaries separating crystalline
regions of different orientations. The temperature at which this transition
occurs remains close to the melting temperature of the vortex lattice in 
samples without columnar pins, if the value of $c$ is 
sufficiently small. As $c$ is
increased while keeping $B$ (and hence the  density of vortex 
lines) constant, the transition temperature begins to increase. This is 
in agreement with theoretical predictions~\cite{YG97}. As $c$ is
increased further, the nature of the transition changes from first-order to
continuous. For small values of $c$, the 
liquid into which the BoG melts has the ``mixed''
character mentioned above -- it is called vortex ``nanoliquid'' in 
Ref.~\onlinecite{Banerjee04}. This partially pinned interstitial 
liquid undergoes a crossover
to a fully depinned vortex liquid as the temperature is increased further.
The temperature at which this crossover occurs is found to be nearly 
independent of the pin concentration, whereas the melting temperature of the
BoG phase increases with increasing $c$. Therefore, the temperature range
over which the interstitial vortex liquid exists becomes narrower with
increasing $c$, and eventually goes to zero. Thus, for sufficiently high 
values of $c$, the BoG melts
directly into a liquid in which all the vortices are delocalized.

Some of the characteristic features found in the experiments 
mentioned above have been reproduced in
simulations of a collection of interacting vortex lines in the presence of
a random array of columnar pins. These include the polycrystalline nature
of the low-temperature BoG phase~\cite{Sen01}, the
increase of its melting
temperature with increasing pin concentration for strong
columnar pins~\cite{TG03}, and the occurrence of an interstitial liquid
phase for low values of $c$~\cite{NH04}. The simulations of 
Ref.~\onlinecite{NH04} have also provided evidence for the existence of an
additional phase, a  
topologically ordered Bragg glass (BrG)~\cite{GLD95}, 
at low temperatures in
systems with a  small concentration of weak columnar pins. 
 
Assuming the
magnetic field to be perpendicular to the layers, the vortex lines in 
highly
anisotropic layered HTSCs may be thought of as stacks 
of pancake vortices
localized on the superconducting layers. These pancake vortices constitute
a system of pointlike objects that interact among themselves and also with
a potential, external to them, arising from the pinning centers. 
We have recently performed a  study~\cite{prlr,prbr} of the 
thermodynamic and structural properties of such a system,
with a very small concentration of randomly 
distributed, strong, columnar pins perpendicular to the layers. 
We used
 a
density functional theory, based on the Ramakrishnan-Yussouff free energy
functional~\cite{ry}, to analyze the 
structure and equilibrium properties of this
classical system of pointlike objects.
In this method
the free energy is expressed as a functional of the time-averaged local
density of the pancake vortices. Different local minima of the free
energy represent different possible phases of the system in this
mean-field description. A numerical minimization procedure 
locates the various minima of the free energy
functional. The phase diagram of the system in the relevant parameter space
was then mapped out by locating the points where the free energies of
different minima cross. 

There are three distinct phases for
very small values ($c \leq 1/64$) of the pin concentration: a
topologically ordered BrG phase~\cite{note}, a polycrystalline 
BoG phase, and an IL phase with coexistence of 
localized and mobile vortices. The BrG is the equilibrium phase  
at low temperatures at these
small values of $c$. As the temperature $T$ is increased, this
phase melts into the IL in {\em two steps} -- the BrG and IL phases are
separated by a small region of the BoG phase. Both the BrG-to-BoG transition
and the BoG-to-IL transition which
occurs at a slightly higher T
were found to be first-order, indicating a two-step first-order melting of 
the nearly crystalline, topologically ordered BrG phase. Experimentally
observed spatial variations of a ``local'' transition 
temperature~\cite{Banerjee03,Soibel01} were also reproduced.

In this paper, we present the results of our investigation of the behavior
of the same system for larger values of the pin concentration, thereby
generating the phase diagram in the
entire relevant area of the $T$-$c$ plane. The methods
used in this study are the same as those used ~\cite{prlr,prbr} 
for the small-$c$ regime. The objective here is
to connect the behavior found for small $c$  with the
well-known results (continuous transition between a low-temperature BoG
phase and a high-temperature depinned vortex liquid) for large $c$. 
Another motivation  is to provide an understanding of some of
the experimental observations mentioned above. 

The main results of our 
present study 
are as follows. We find that the width (in temperature) of the intermediate
BoG phase that separates the low-temperature BrG and high-temperature IL
phases increases initially as the pin concentration $c$ is increased from a
very small value at fixed $B$. 
As $c$ is increased further, the BrG minimum of the free
energy ceases to be the one with the lowest free energy at low temperatures.
This observation indicates that the BrG phase is thermodynamically
stable only if the pin concentration is smaller than a critical value. This
critical value of $c$ turns out to be close to 1/32 for the system 
parameters (appropriate for BSCCO) used in our calculation, at $B= 2$kG. 
For higher values
of $c$, we find a low-temperature polycrystalline BoG phase and a
high-temperature IL phase. The transition between these two phases becomes
continuous as the value of $c$ is increased. 
A continuous transition is difficult to
locate in a  mean-field calculation. 
In that case the minimum representing the
low-temperature BoG phase transforms continuously into  that corresponding
to the high-temperature IL phase as the temperature is increased: there is
no free-energy crossing that can be used to define the transition temperature.
Indeed, it becomes difficult to distinguish
a true 
continuous transition from a crossover. Guided by existing theoretical
and experimental results (summarized above), we assume that there is a 
continuous transition, and then locate the transition temperature using a
criterion based on {\it the occurrence of percolation of liquid-like regions 
across the sample}. The transition temperature
obtained from this definition was found~\cite{prlr,prbr,prbv} 
at lower $c$ to coincide with that obtained from free energy
crossings.
The details are explained in Sec.~\ref{res} below.
In a study~\cite{Shes95} 
of the two-dimensional
disordered bosonic Hubbard model, it was found that the transition between the 
BoG and superfluid phases nearly coincides 
with the onset of percolation of sites that have large values of the
superfluid order parameter. Since the system we are considering can be
mapped approximately to a two-dimensional system of interacting bosons in 
a random potential, we believe that the percolation-based criterion we have
used to determine the transition temperature is quite reasonable.

The BoG--IL transition temperature obtained this way increases
 slightly
as the pin concentration is increased at fixed $B$. The BoG phase continues
to exhibit a polycrystalline structure as $c$ is increased: the typical
size of crystalline domains decreases with increasing $c$. To estimate
the depinning temperature, we monitor the average value of the integrated
density of pancake vortices inside the range of the pinning potential of 
each columnar defect. This quantity decreases as the temperature is increased.
We define the depinning temperature as that at which the
integrated density falls below an appropriately chosen critical value. 
The depinning temperature obtained this way decreases slowly with
increasing $c$. This happens because~\cite{prbv,daf} two pinning
centers can not be simultaneously occupied by vortices if they happen to be
very close to each other. This effect reduces the average pinning efficiency
of a defect as $c$ is increased. For the range of $c$
we have considered ($c \leq 1/8$), the depinning temperature is found to
lie above the temperature at which the BoG phase transforms into a liquid. 
So, for these values of the
pin concentration, the liquid phase into which the BoG melts can be 
classified  as IL, with a coexistence of pinned and delocalized vortices.
Our results indicate that the IL would disappear (the BoG would melt into
a depinned liquid) at a  value of $c$ just slightly
higher than $1/8$. From that point on, there would be a single transition
between the low-temperature BoG and a high-temperature depinned vortex 
liquid: thus our work encompasses
all the interesting parts of the phase diagram in the $c$ vs. $T$ plane.
Using a criterion
based on the degree of localization of the vortices, we define a local
transition temperature and find that this temperature exhibits spatial
variations similar to those found in experiments~\cite{Banerjee03,Soibel01}.
   
The rest of this paper is organized as follows. In section~\ref{meth}, we
describe the model and the computational method used in our study. The
main results obtained in this work are presented and discussed in detail in
section~\ref{res}. The last section~\ref{con} contains a summary of our
main results and a few concluding remarks.

\section{Method\label{meth}}

The general method used in this study has been 
described previously\cite{prbv,prbr}
and we will keep the discussion here to the minimum
necessary to establish notation. 
We consider the limit of highly anisotropic layered
superconductors with vanishing interlayer Josephson coupling. In this limit,
the energy of a system of pancake vortices can be written as a sum over
anisotropic pair-wise interactions.
This limit
is appropriate to HTSCs of the Tl and Bi families.
We will use material parameters appropriate to BSCCO throughout
this work. 

The pinning potential due to a collection of columnar pins perpendicular to 
the layers is the same on all the layers. This implies that the time-averaged
local density $\rho({\bf r})$ ($\bf r$ is a two-dimensional vector 
denoting the position on a layer) 
of pancake vortices is also the same on all the layers. This makes the problem
effectively two-dimensional~\cite{prbv,prbr}.
We conduct a numerical
search for local and global minima of the free energy $F$ written
as a functional of 
$\rho({\bf r})$:
\begin{equation}
F[\rho]-F_0=F_{RY}[\rho]+F_p[\rho] .
\label{fe}
\end{equation}
Here $F_0$ denotes the free energy of a uniform
vortex liquid of density $\rho_0=B/\Phi_0$, with $B$ being
the magnetic induction and
$\Phi_0$  the superconducting flux quantum.
In the absence of pinning, the free energy functional
is given by the Ramakrishnan-Yussouff\cite{ry} expression:

\begin{eqnarray}
\beta F_{RY}[\rho] &=& \int d^2r\left[\rho({\bf r})
\{ \ln(\rho({\bf r}))-\ln (\rho_0)\} -
\delta\rho({\bf r})\right] \nonumber \\ 
&-&\frac{1}{2}\int d^2r \int d^2r^\prime \,\tilde{C}(|{\bf r} -
{\bf r}^\prime |)
\delta\rho({\bf r}) \delta\rho({\bf r}^\prime), 
\label{ryfe}
\end{eqnarray}
where $\beta$ is the inverse temperature, 
$\delta\rho({\bf r}) \equiv \rho({\bf r}) - \rho_0$ the deviation
of the density from the uniform value,
and $\tilde{C}(r)$ a direct
pair correlation function\cite{han}, for which we use the
hypernetted chain\cite{menon} results. 
With
these assumptions, the results depend on the in-plane London
penetration depth $\lambda(T)$ and on the dimensionless parameter $\Gamma$
given by:

\begin{equation}
\Gamma = \beta d \Phi^2_0/8 \pi^2 \lambda^2(T).
\label{gamma}
\end{equation}
We take $d=15 \AA$ for the interlayer distance in BSSCO, and
the standard two-fluid form for $\lambda(T)$
with $\lambda(0)= 1500 \AA$ and $T_c=85$K (at zero field).

The remaining term in the right-hand side of Eq.(~\ref{fe}) represents
the pinning and is of the form:

\begin{equation}
F_p[\rho] = \int d^2r V_p({\bf r}) \delta\rho({\bf r}). 
\label{pin}
\end{equation}
where the pinning potential $V_p$ is written as a sum extending
over the positions of the pinning columns in the transverse plane:
\begin{equation}
V_p({\bf r})=\sum_j V_0(|{\bf r}-{\bf R}_j|).
\label {pinpot}
\end{equation}
The potential $V_0$ corresponding to a single pinning center
is chosen to be of the usual truncated parabolic form~\cite{daf}
\begin{equation}
\beta V_0(r)=-\alpha\Gamma[1-(r/r_0)^2]\Theta(r_0-r).
\label{single}
\end{equation}
We use the quantity $a_0$ defined by $\pi\rho_0a_0^2=1$ as our
unit of length. In terms of this quantity, we take 
$r_0=0.1 a_0$. As for the dimensionless strength $\alpha$, we 
use $\alpha=0.05$. For this combination of values\cite{prbv}
each pinning center traps nearly one vortex in the lower
part of the temperature range studied.

The minima of the free energy are found by discretizing the density
over a computational triangular lattice of spacing $h$. The value of $h$ 
is chosen to be sufficiently small to provide an accurate description of the
spatial variations of $\rho({\bf r})$. The set of discretized variables
$\{\rho_i\}$ is given by $\rho_i\equiv \rho({\bf r}_i)\sqrt{3}h^2/2$,
where $\{{\bf r}_i\}$ represents the set of 
computational lattice points. 
One then minimizes the free energy with respect to the  $\{\rho_i\}$
set by using a procedure\cite{cdo} that ensures that the nonnegativity
constraint on these variables is satisfied. 

\section{Results\label{res}}

Using the methods described above, we have studied relative pin 
concentrations
 $c$ ($c=N_p/N_v$ as defined above) 
ranging from $c=1/32$ to $c=1/8$, keeping  $B$ fixed at 2 kG. 
Results for smaller concentrations have
been  previously\cite{prlr,prbr} reported.  All results reported here are for a 
computational lattice size $N=1024$ and the  computational lattice
spacing is $h=a/16$ where $a=1.998 a_0$ is\cite{prbv}
the equilibrium lattice constant
of the Abrikosov vortex lattice in the absence of pinning 
in the temperature range considered (see below)
which is chiefly determined by the melting temperature of the
unpinned vortex lattice, namely $T_m^0 \simeq $ 18.4K  
at the field studied. For these values of  $h$ and $N$ the
system contains $N_v=4096$ vortices and therefore
the number of pins ranges
from $N_p=128$ to $N_p=512$. These pins are set at random positions. The
results are very consistent from sample to sample and it turns out
to be sufficient to average over five such samples.

The initial conditions used for the variables $\rho_i$ to
start the minimization process are of
two kinds: the first kind is ``liquid-like'' with uniform density, and
the second is a ``crystal-like'' initial state with values of the
$\rho_i$ variables obtained, for a given random pin configuration,  by 
minimizing the pinning energy with respect to all the possible 
crystal configurations related by symmetry operations of the triangular
lattice. The first kind of initial conditions leads, if one quenches
to a sufficiently high temperature (higher than $T_m^0$),  to free
energy minima which correspond to a liquid state, as revealed
by the methods of analysis discussed below. The second
kind leads\cite{prbr} to a BrG state if quenched to a sufficiently low
$T$ provided (as we shall see) that $c$ is not too high. BoG states
can be obtained by slowly (we typically use $0.2$K temperature increments)
cooling liquid states obtained as described above, or by
quenching to sufficiently low $T$
either with the first kind of initial conditions  or with the second
kind at higher $c$, as we shall see below.

\begin{figure}
\includegraphics [scale=0.5] {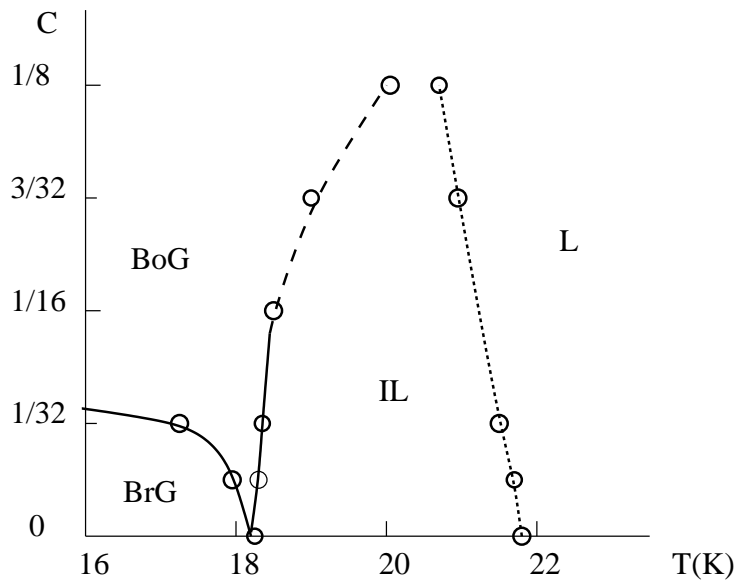}
\caption{\label{fig1} The temperature, $T$, vs relative pin 
concentration, $c$, phase
diagram obtained from our results for $B$ = 2kG. The labels L,
IL, BoG and BrG denote liquid, interstitial liquid,
Bose glass and Bragg glass phases, respectively. 
Open circles represent actual data points, and the
lines are guides to the eye. The solid lines indicate
first-order transitions, the dashed line, continuous transitions, while
the dotted line marks the location of the depinning crossover. See text for
details.}
\end{figure}

From the set of $\{\rho_i\}$ values obtained at each minimum 
one can  calculate the density
correlations, such as the structure factor $S({\bf k})$:
\begin{equation}
S({\bf k})=|\rho({\bf k})|^2/N_v
\label{sofk}
\end{equation}
where $\rho({\bf k})$ is the discrete Fourier 
transform of the  $\{\rho_i\}$. 
Equivalently, one
can consider the real-space correlations
through the 
discrete Fourier transform of $S({\bf k})$,
which is the two-point
spatial correlation function $g({\bf r})$ of the time-averaged local
density. It is also useful to consider the angularly averaged 
counterparts of these two quantities, which we will call
$S(k)$ and $g(r)$ respectively.

However, evaluating this
and other correlations is not all one can do: one can use the set $\{\rho_i\}$ 
to directly visualize the density distribution. Further, one can examine the
structure formed by the {\it vortices} themselves. To do so, one extracts from
the $\{\rho_i\}$ variables the local peak densities: we define site $i$ as
being a local peak location  if $\rho_i$ is larger than any value of $\rho_j$
evaluated at a site $j$ within a radius $a/2$ from $i$, where $a$ is the
lattice constant of the  vortex lattice in the absence of pinning. We refer
to the network formed by these peak locations as the ``vortex lattice''. At
low temperatures, the number of peaks matches the number of vortices $N_v$,
indicating that, as expected, the vortices are localized. This ``vortex
lattice'' can be studied by direct visualization, and also by carrying
out a Voronoi construction for it, that is, by examining its defect
structure.

\begin{figure}
\includegraphics [scale=0.4] {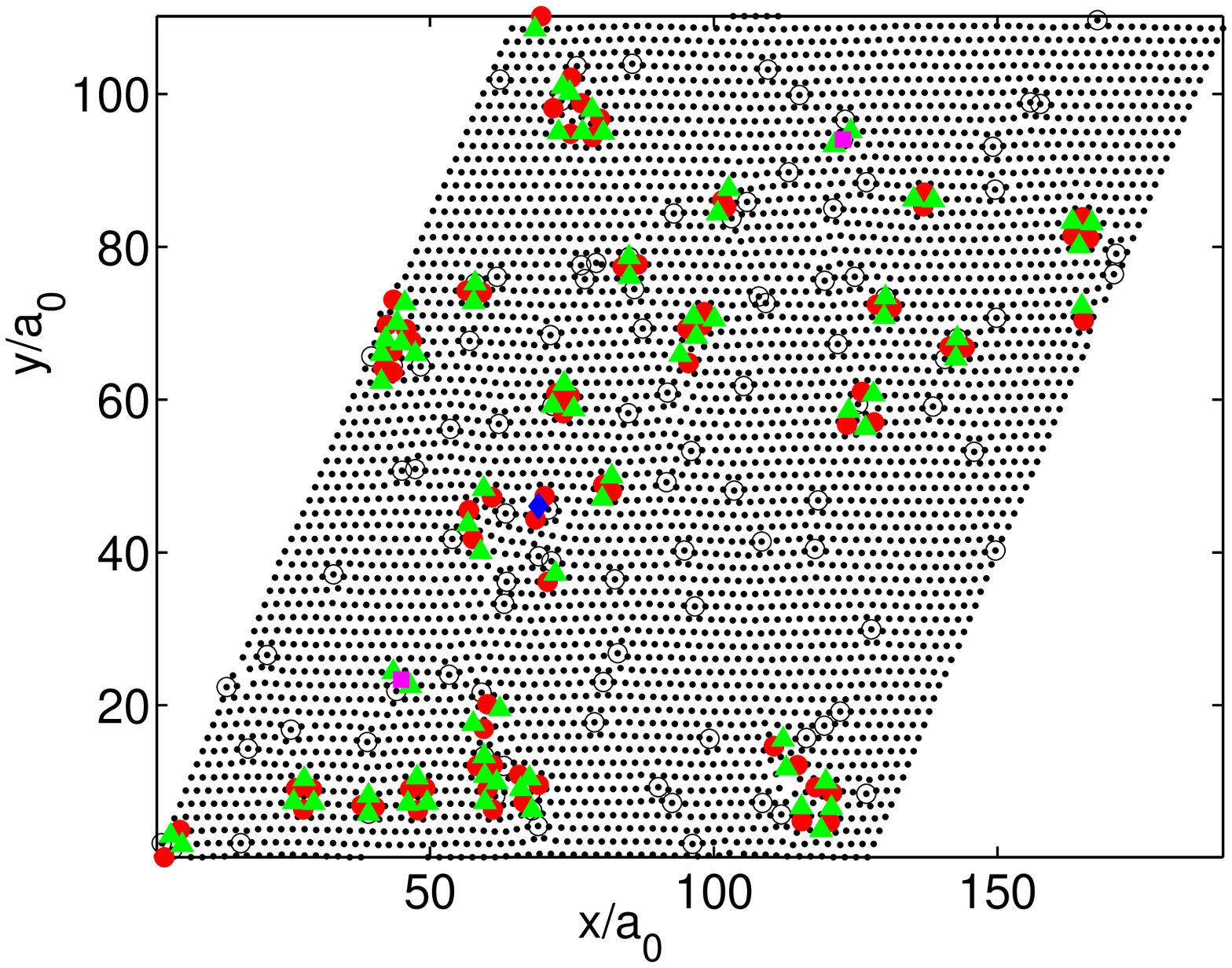}
\includegraphics [scale=0.4] {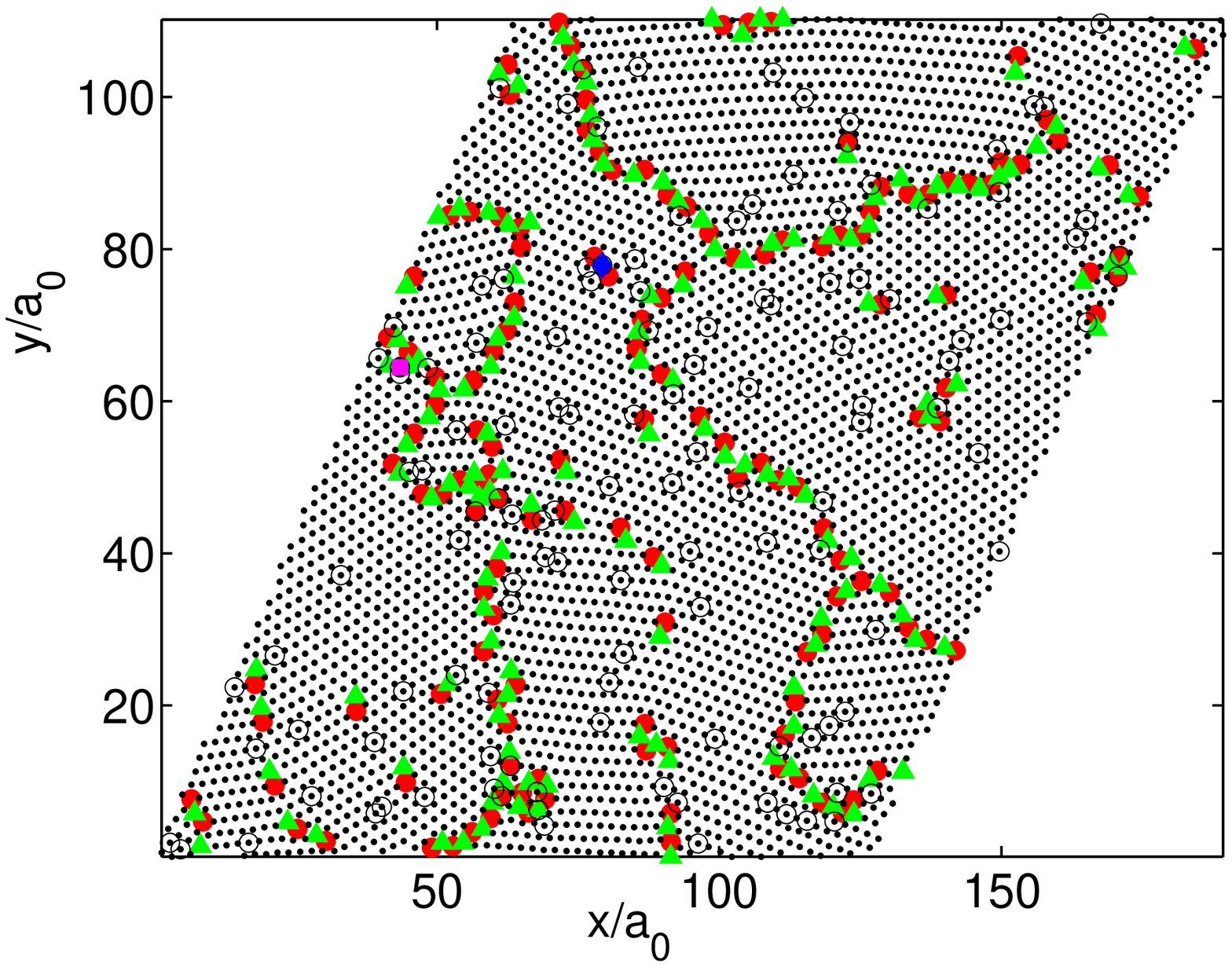} 
\caption {\label{fig2} (Color online) Voronoi plots for the BrG (left)
and BoG (right) states found at $T=17$K for one of the samples with
$c=1/32$. The black dots denote sites with six neighbors, the 
(red)  solid circles label fivefold coordinated sites, and the (green)
triangles sevenfold  coordinated sites. A few occurrences of fourfold 
and eightfold coordinated sites are indicated by (magenta) squares and 
(blue) diamonds, respectively. Sites surrounded by black  
circles are where pinning centers are located. }
\end{figure}

To describe orientational order, we  will also study the bond-orientational
correlation function $g_6(r)$ in the vortex lattice which we define as:
\begin{subequations}
\label{angular}
\begin{equation}
g_6(r)=\langle\psi({\bf r})\psi(0)\rangle
\label{ang1}
\end{equation}
where  $\langle\cdots\rangle$ brackets denote  average over the vortex
lattice and the field $\psi({\bf r})$ is:
\begin{equation}
\psi({\bf r})=\frac{1}{n_n} \sum_{j=1}^{n_n} \exp[6i\theta_j({\bf r})]
\end{equation}
\end{subequations}
where $\theta_j({\bf r})$ is the angle made by the bond connecting
a vortex at ${\bf r}$ to its $j$-th neighbor and a fixed
axis, and $n_n$ is the number of neighbors of the vortex at ${\bf r}$, as
determined from a Voronoi construction.

\begin{figure}
\includegraphics[scale=0.4] {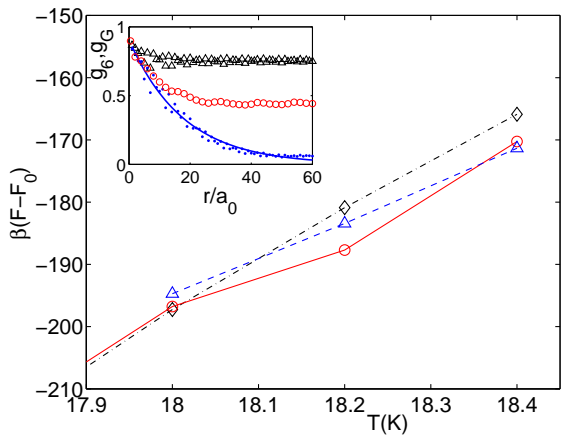}
\caption{\label{fig3} (Color online) The main plot shows the free energies
as a function of $T$ for three states obtained for the same
sample as in Fig.~\ref{fig2}, displaying
the first-order transitions at the crossings.
The symbols represent the data points, joined
by straight segments. (Black) diamonds and dash-dotted line: BrG. (Red)
circles and solid line: BoG. (Blue) triangles and dashed line: IL. In
the inset, several correlation functions
are plotted versus dimensionless distance, all at $T=17$K and averaged over
five configurations: the correlation function $g_6(r)$ (Eq.~(\ref{ang1})) is 
shown
for the BrG  ((black) triangles) and the BoG
((blue) dots) states,  with the solid curve being an exponential 
fit to the dots. The (red) open circles represent the data for the function 
$g_G(r)$ (Eq.~(\ref{bo}))
for the BrG state.}
\end{figure} 

Quite analogously, one can  define 
a ``translational correlation function'' $g_G(r)$ of the vortex lattice 
by an equation identical
to the right-hand side of Eq.~(\ref{ang1}), but with the field $\psi$
replaced by
\begin{equation} 
\psi_G({\bf r})=\exp(i {\bf G} \cdot {\bf r})
\label{bo}
\end{equation}
where ${\bf G}$ is one of the shortest nonzero
reciprocal lattice vectors of the triangular vortex lattice
in the absence of pinning. We will average over the  six equivalent $\bf G$'s. 

Different free energy minima at
a given
temperature are found, as in previous work\cite{prlr,prbr}
by starting the minimization  program with different initial conditions,
either in a  quenching mode, where the initial conditions are liquid-like
or crystal-like,
or on a warm-up (or cool-down) mode 
where the density values for
a certain minimum are used as input to find a similar minimum at a 
slightly higher 
(or lower) temperature.
Determination of the nature of a given 
free energy minimum is carried out by
careful consideration of all the information obtained, both in the original
computational lattice and on the vortex lattice.

\begin{figure}
\includegraphics [scale=0.65] {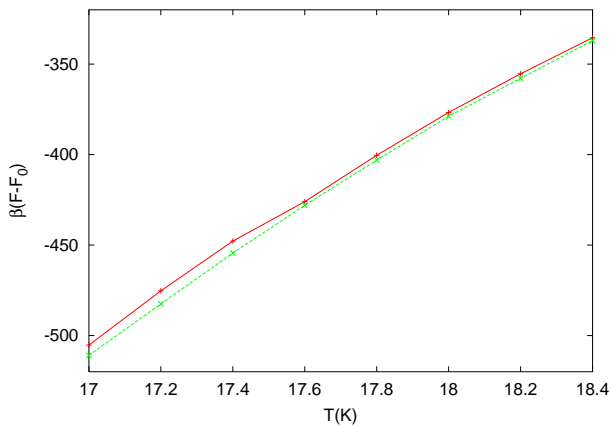}
\caption{\label{fig4}(Color online) Free energy (in units of $k_BT$)
of two different states at a concentration $c=1/16$,
as a function of temperature. The (red) plus signs 
and solid line correspond to a state obtained by quenching
to $T=17$K with
crystalline initial
conditions and then slowly warming. At concentrations $c\le 1/32$
this produces a BrG state. The (green) crosses and dashed line 
correspond to a state similarly obtained but by starting with uniform
initial conditions. This state is BoG-like at all $c$ values. One
can see that the attempted BrG state has, at this
concentration, a higher free energy, with
the difference actually increasing at low $T$.}
\end{figure}

\subsection{Phase diagram}\label{phdia}

We start the discussion of our results by summarizing the phase diagram
we obtain in Fig.~\ref{fig1}. The data which lead to this diagram and
results supporting it
will be discussed in the rest of this section. The results shown are for a 
fixed value of the magnetic induction, $B$ = 2kG. We see in Fig.~\ref{fig1}
that at small concentrations $c$ of random pins, and low $T$, one has
a nearly crystalline, topologically ordered BrG phase which, upon
warming, melts into an IL in two 
first-order steps, indicated by solid lines. As already
seen in Refs.~\onlinecite{prlr} and \onlinecite{prbr} for $c\le 1/64$, 
the low-$T$ BrG
phase and the higher
$T$ IL are separated by a sliver of polycrystalline 
BoG phase. We find here that, as
one increases $c$, the temperature region where the intermediate BoG occurs
first becomes wider. However, as $c$ exceeds 1/32, the BrG drops out of
the picture: the free energy of any phase that may be identified
as BrG-like becomes, as  we shall see, too high, and the only transition
is between the IL and a BoG phase. This transition is, furthermore,
not first-order but continuous (dashed line).  
The transition line bends markedly
towards higher temperatures as $c$ increases. We have indicated also in
the figure (dotted line) the depinning crossover, determined as explained
below, which indicates the temperature at which the pins are no longer
able to effectively pin the vortices. This line, which separates
the IL from the ordinary vortex liquid, is weakly $T$ dependent, and very near 
the highest $c$ studied, it merges with the IL  to BoG line, indicating that,
from that point on, the phase diagram becomes very simple, as only a single
transition between an unpinned liquid phase and a pinned BoG
phase occurs.

\subsection{Results for $c=1/32$}\label{smallc}

We begin by considering the concentration $c=1/32$. All of our results are
obtained on a computational lattice of size $N=1024$ so that, with the value
of $h$ chosen as discussed above, we always have $N_v=4096$ vortices. Therefore
at this concentration we have $N_p=128$ pins. We can obtain at this
concentration three kinds of states: the first kind is obtained by quenching
the system to a rather low $T$ with crystalline initial conditions. An example
(discussed in more detail below) is shown in the left panel of Fig.~\ref{fig2}
which depicts the vortex lattice structure (positions of local density peaks).
The second kind is obtained by quenching to similar temperatures with
liquid-like initial conditions. An example is shown in the right panel of
the same figure. Finally, if one quenches with liquid initial conditions to
a high temperature (e.g. $T=18.6$K), one obtains a third state which is
very obviously liquid-like. This is  as previously\cite{prbr}
found  for $c=1/64$.

We show in Fig.~\ref{fig2} Voronoi plots for
the vortex lattice at $T=17.0$K. These plots show  
the defect structure of the two different kinds of low temperature 
states. One can see  in the 
left plot  a nearly crystalline structure, with defects
localized and consisting of paired dislocations and larger clusters
of tightly bound dislocations with net Burgers vector equal to zero. 
The second plot
is quite different: there are many more defects and they are not randomly
distributed, but they are organized forming continuous grain boundaries
which give the sample an overall ``polycrystalline'' structure. Therefore,
one is led to identify the first state as the BrG, and the second, as the BoG.
Examination of the correlation functions confirms that this situation
is in all ways identical to that previously found\cite{prbr} at smaller
values of $c$. For this reason, 
many of the details will not be repeated here. 

\begin{figure}
\includegraphics [scale=0.65] {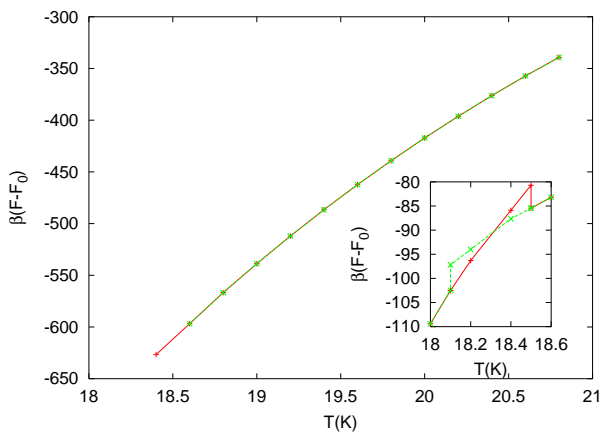}
\caption{\label{fig5}(Color online) Main figure: The dimensionless
free energy for a concentration  $c=1/8$. The (red) plus signs
and (green) crosses represent results obtained on warming and cooling
respectively. They are so similar that the symbols merge into an
apparent asterisk. In the inset the results of a similar cycling at
$c=1/64$ are shown, with the same symbols: clearly now a hysteretic
first-order transition occurs. }
\end{figure}

The free energy analysis for this sample is shown in the main plot
of Fig.~\ref{fig3}. The data for the BoG and BrG states
were obtained by warming up the states shown in Fig.~\ref{fig2},
while the IL state data were obtained from cooling the
high-$T$ liquid. This IL state becomes unstable if one attempts
to cool it further; similarly, the two solid states found at low temperatures
eventually melt if warmed up enough. 
Here one can see that the IL state transforms, upon lowering $T$, to the
BoG state via  a first-order transition, while, upon further cooling,
there is a second transition to the BrG. For this sample, the 
spread between the two transitions is about $0.4$K but this number is rather
variable from sample to sample. The average is represented in Fig.~\ref{fig1}.
There is a very clear increase in the width of this interval, compared to that
for $c=1/64$.

The results shown in the inset in Fig.~\ref{fig3}  
confirm our identification of the low-$T$ phases.
Here, different correlation functions
are plotted versus  distance measured in units of $a_0$.
These are for $T=17$K and represent averages over
five configurations. For the BrG state
we plot the bond-orientational correlation 
function $g_6(r)$ (Eq.~(\ref{ang1})),  
[(black) triangles] and also 
the translational correlation function
$g_G(r)$ defined in Eq.~(\ref{bo}) [(red) open
circles]. It can be seen that $g_6(r)$ exhibits long-range
order, and that the decay of $g_G(r)$ at long distances is much slower
than the initial exponential decay. 
The remaining data points [(blue) dots] show $g_6(r)$ for the BoG state. 
In this case, there is no long-range order: the solid curve is an exponential
fit to the data. These results are quite consistent with the 
defect structures found in the Voronoi plots.

\begin{figure}
\includegraphics [scale=0.65] {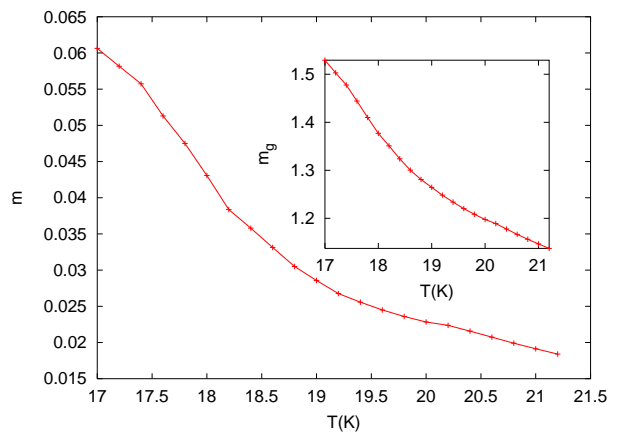}
\caption{\label{fig6}(Color online) Degree of translational order,
measured in two different ways, as a function
of temperature at $c=3/32$. In the
main plot, the ``order parameter'' $m$ obtained from the structure 
factor (see Eq.(\ref{mdef}))
is plotted, while
in the inset the height of the first peak at finite $r$ 
in $g(r)$ is displayed. 
In all cases the symbols
are data points joined by straight segments.}
\end{figure}

\subsection{Results for $c>1/32$}\label{largec}

The behavior described above changes, however, when $c$ is 
increased above the value
of 1/32. In Fig.~\ref{fig4} we  have shown some free energy results
(in units of $k_BT$)
obtained at $c=1/16$. Results are shown, for the same pin configuration, for
two states both obtained  by quenching to $T=17.0$K and then warming up.
One of the states [(red) plus signs and solid line] is obtained
from crystal-like initial conditions, while for
the other set of data [(green) crosses
and dashed line], the initial conditions are liquid-like. At smaller
$c$, the state obtained from crystal-like initial conditions 
is BrG-like and is the lowest free-energy state 
at low $T$, while the other state is a BoG, and is the lowest free-energy
state at temperatures close to melting.  This is not the case here: the
state obtained from liquid-like initial conditions
is in fact a BoG, but the  
putative BrG (which is rather more disordered than the BrG 
states found for
smaller values of $c$) turns
out to have a higher free energy than the BoG, and the difference
becomes larger at lower $T$. 
There is therefore no longer a transition
between the BoG and BrG states. At higher values of $c$, the same situation 
obtains, except that the difference in free energies between the states
obtained using the two different initial conditions becomes larger. 
Also, the minima obtained from crystal-like
initial conditions begin to exhibit unpaired dislocations, so that they can
no longer be convincingly identified as BrG.
This indicates that the BoG to BrG first-order 
transition line  turns over near $c=1/32$ as indicated in the phase diagram
of Fig.~\ref{fig1}.

We will therefore from now on ignore the BrG-like state at these larger values
of $c$, since it is never the equilibrium state.
The nature of the remaining transition, from IL to BoG, also changes in
this concentration range: it becomes a continuous transition. This can be
seen in the
main part of Fig.~\ref{fig5}, where we plot the dimensionless free energy as
a function of $T$ for a pin configuration with $c=1/8$. The BoG state obtained
in the usual way by quenching to low $T$ is warmed up, [(red) plus signs and
solid line]  and then cooled down again [(green) crosses, dashed line]. The
values of the free energy obtained in the heating and cooling runs are
the same within numerical error: this is 
made evident by the symbols merging into each other and forming (bi-colored)
asterisks.  This behavior is in contrast with that found at lower $c$, as
exemplified in the inset of Fig.~\ref{fig5}, which is for a system
subjected to similar temperature cycling at $c=1/64$. In that case one can see
the obvious hysteresis associated with the first-order transition
between BoG and IL (the transition to BrG at a lower temperature
is not shown in this plot). Plots
similar to those in Figs.~\ref{fig4} and the main plot
of Fig.~\ref{fig5} can be obtained for any pin configuration with 
$c \ge 1/16$. 
Thus, there is only one transition in this higher concentration range
and it is, furthermore, continuous.  Further evidence for this, and
more details on this transition are shown in the next few figures.

\begin{figure}
\includegraphics[scale=0.4] {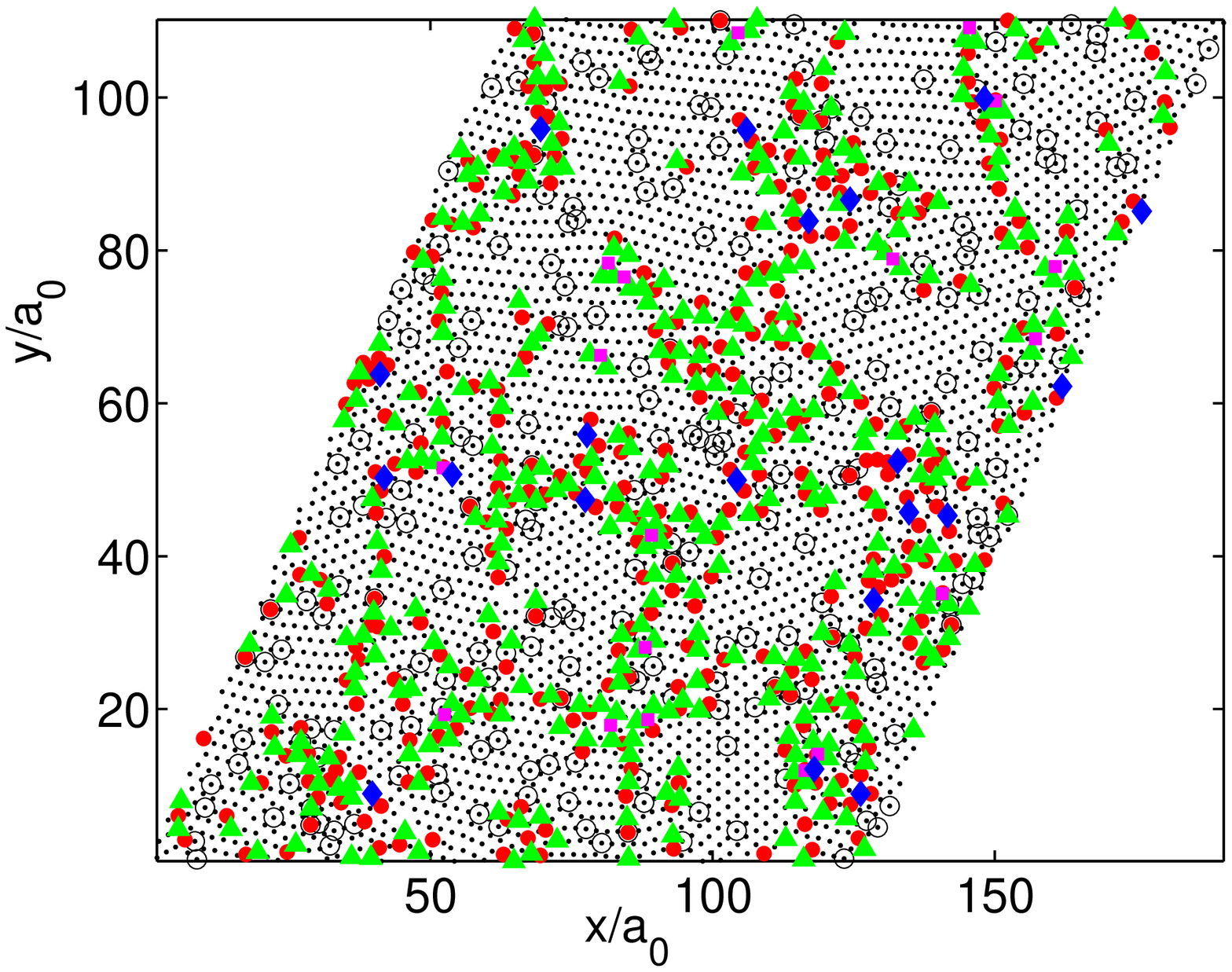}
\includegraphics[scale=0.4] {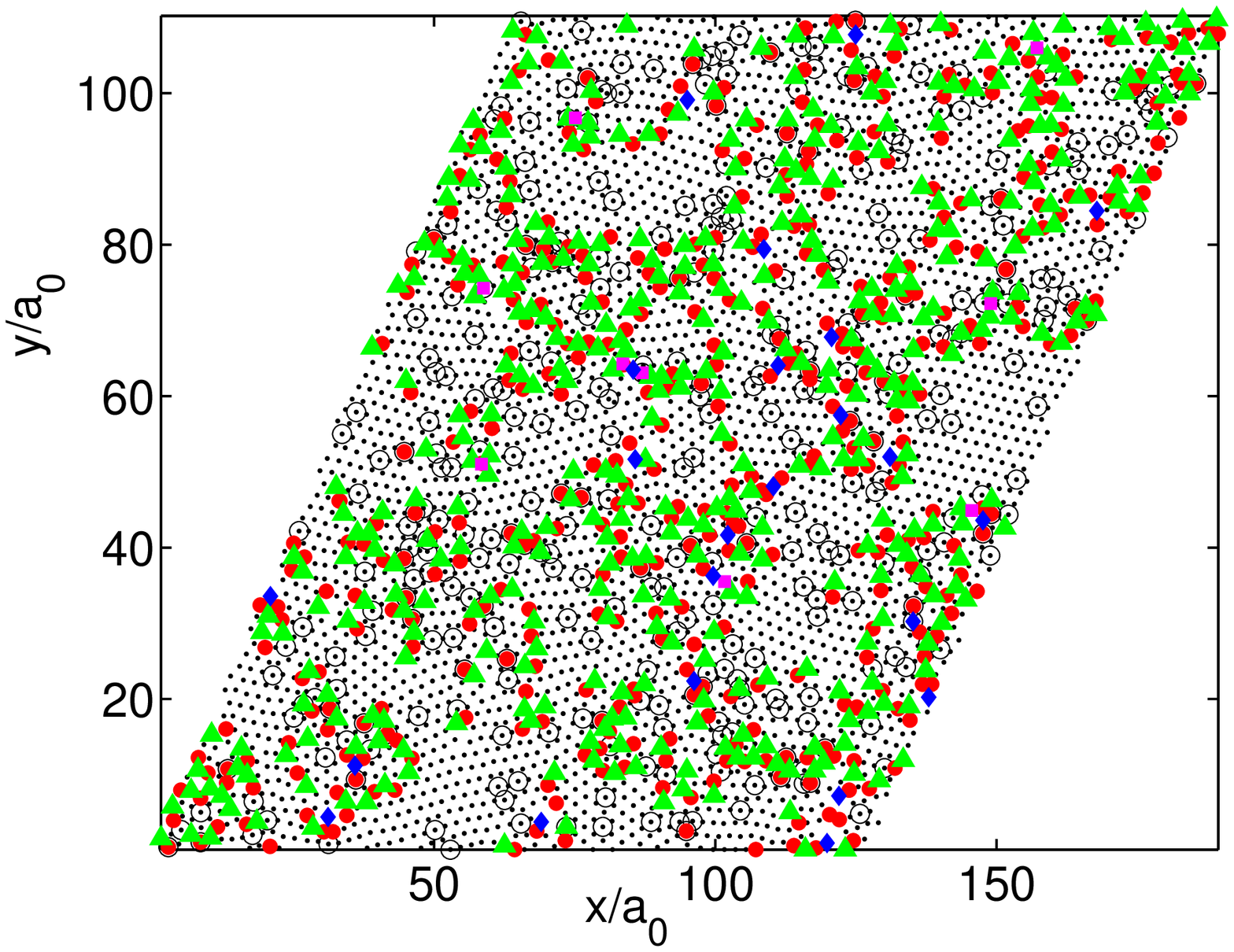}
\includegraphics[scale=0.4] {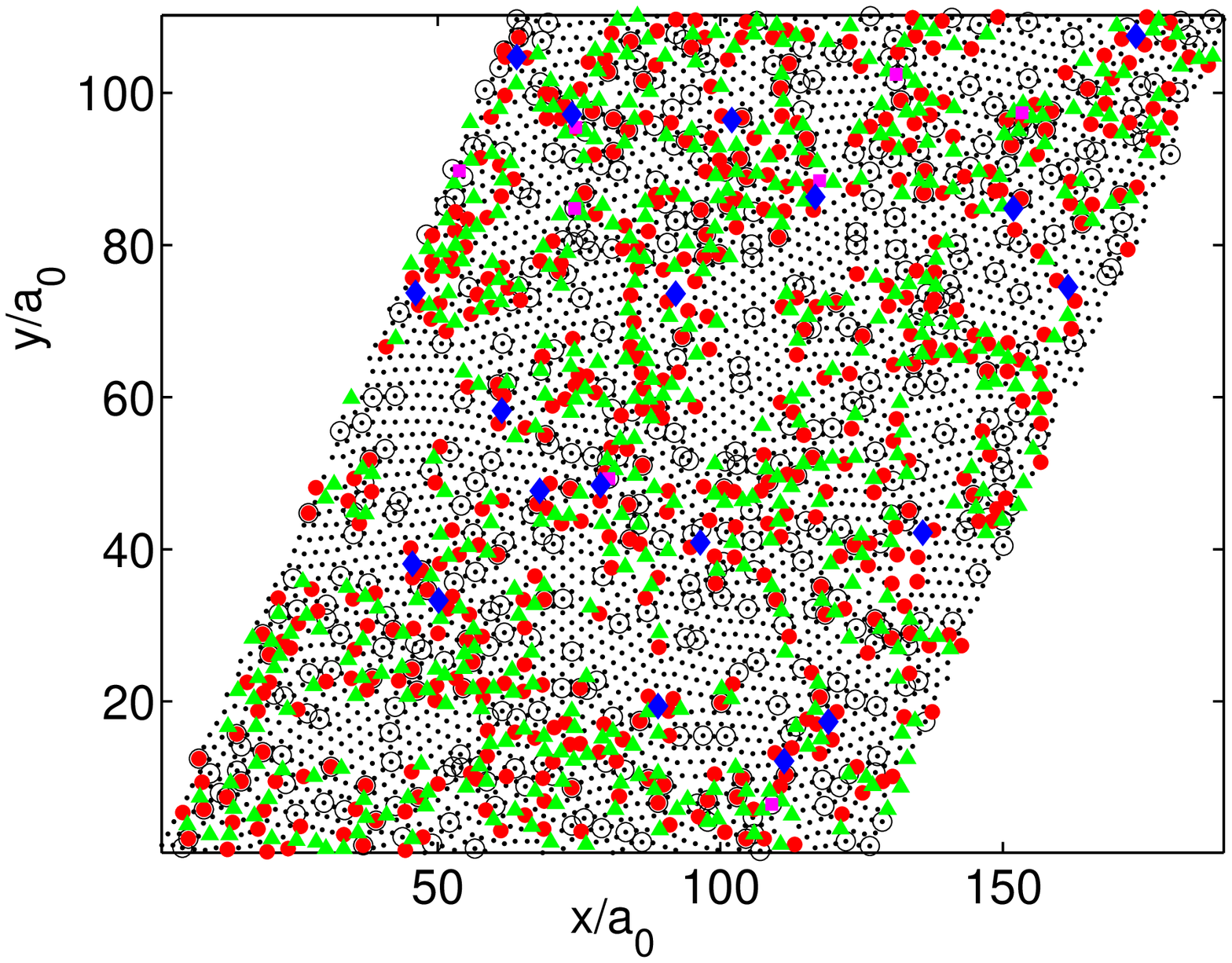}
\caption{\label{fig7}(Color online) Voronoi plots for samples
with $c=1/16$, $c=3/32$, and $c=1/8$ (from top left to bottom) at
$T=18.4$K. The meaning of the symbols is the same as in Fig.~\ref{fig2}.}
\end{figure} 

In Fig.~\ref{fig6} we give an example of the behavior of the degree of
translational order in the system, 
as a function of $T$, at a concentration $c=3/32$. Two different
measures of the degree of translational order are plotted. One, displayed in
the main plot, is
the value of an ``order parameter'' $m$ obtained from the structure factor
$S({\bf k})$. It is defined as
\begin{equation}
m \equiv \sqrt{S_{max}/N_v},\label{mdef}
\end{equation}
where $S_{max}$ is the largest value of $S({\bf k})$ averaged over the six 
$\bf k$ vectors related to one another by lattice symmetry. 
This is the more traditional measure of the degree of translational order:
$m$ would be equal to unity for a perfect lattice.
The data points represent averages over five configurations and
are obtained by initially quenching the system to $T=17$K with liquid-like
initial conditions, and then warming up. Clearly, no sharp change in the 
value of the quantity plotted occurs, even though it changes by a factor
of three in the $T$ range considered. In the
inset, we plot the value $m_g$ of the first maximum at finite $r$ of
the angularly averaged correlation function $g(r)$. The behavior
is qualitatively similar, and again shows no evidence of any discontinuity,
even beyond the temperature where the system is expected to melt and
where, as we shall see below, the system is in fact in an IL state.  This
is in contrast with the discontinuous changes in the order parameter found upon
melting at lower $c$ (see e.g. Fig.~6 in Ref.~\onlinecite{prbr}). The plots
in Fig.~\ref{fig6} exhibit largest slopes near $T \sim 18.5$K, suggesting that a
continuous transition (or a crossover) takes place near this temperature.

We next visualize the phases involved as a function of $T$
and $c$ through Voronoi and peak-density
plots for the vortex lattice. As explained above, we consider only the
states obtained from liquid-like initial conditions. As shown in
Fig.~\ref{fig5}, these states can be
warmed up and cooled down reversibly.
In Fig.~\ref{fig7} we show Voronoi plots, all at the same 
temperature $T=18.4$K, 
for three concentrations, ranging from 1/16 to 1/8.
The temperature chosen is slightly to
the left of the dashed transition line in Fig.~\ref{fig1}: the plots
still show BoG character.  As one can see, the
nature of the state does not change with pin concentration, except
that the number of defects obviously increases. The crystalline 
domains are still quite well defined, but they
become smaller as $c$ increases. 
Our results for the dependence of
the number of defects on $c$ are consistent with this number
being proportional to $\sqrt{c}$.
This implies that
the number of crystalline domains in the BoG state 
is proportional to 
$c$. This being so, the area of a typical crystalline grain 
would be proportional to $1/c$, and the length of
its perimeter  to $1/\sqrt{c}$. Hence, the total
length of the grain boundaries in the sample 
(which is approximately proportional to the number of 
dislocations since nearly all the dislocations appear at grain boundaries)
would be proportional to $\sqrt{c}$. This is similar to the experimental
results reported in Ref.~\onlinecite{Menghini03}.

This result suggests a simple qualitative explanation for the appearance of 
the BrG phase at very small values of $c$ and its 
subsequent disappearance. The elastic energy cost for the 
formation of the grain boundaries in the BoG phase should be
proportional to the total length of the grain boundaries, i.e. to $\sqrt{c}$,
neglecting the contribution from  interactions between grain boundaries
for small $c$. The pinning energy is minimized in the BoG phase by having
essentially all pinning centers  occupied by vortices in this phase
(see section \ref{depin} below). The free energy cost
of forming the BrG state arises from two sources: First its pinning energy is
higher because pinning centers that lie far away from any
lattice point of the crystalline initial state used for obtaining a BrG
minimum are not fully occupied by vortices (a similar behavior was found in the
simulation of Ref.\onlinecite{NH04}). Second, 
there is  extra elastic energy associated with the formation of the defect
clusters with zero net Burgers vector. These defect clusters, found 
mostly near pinning centers, are formed as 
the effect of the randomly located pinning centers is accommodated in the
initial crystalline structure by small displacements of the 
vortices from their ideal
lattice positions towards the nearest pinning center. These ``uncharged''
defect clusters, called ``twisted bond defects''~\cite{leiber}, cost relatively
little energy because they can be formed by distorting the lattice at short
range. The energy cost of forming these defect clusters is proportional
to their number at the relevant small $c$ values. Our numerical results 
for the 
$c$-dependence of the number of unoccupied (or partially occupied)
pinning centers and the number of
dislocations in the BrG state suggest that the
energy cost arising from both these sources is 
proportional to the defect
concentration $c$. Since $\sqrt{c}$ grows faster than $c$ as $c$ is 
increased from zero, the BrG phase would be favored over the BoG phase at
very small values of $c$, but not at higher ones, as found in our study.
  
\begin{figure}
\includegraphics[scale=0.4] {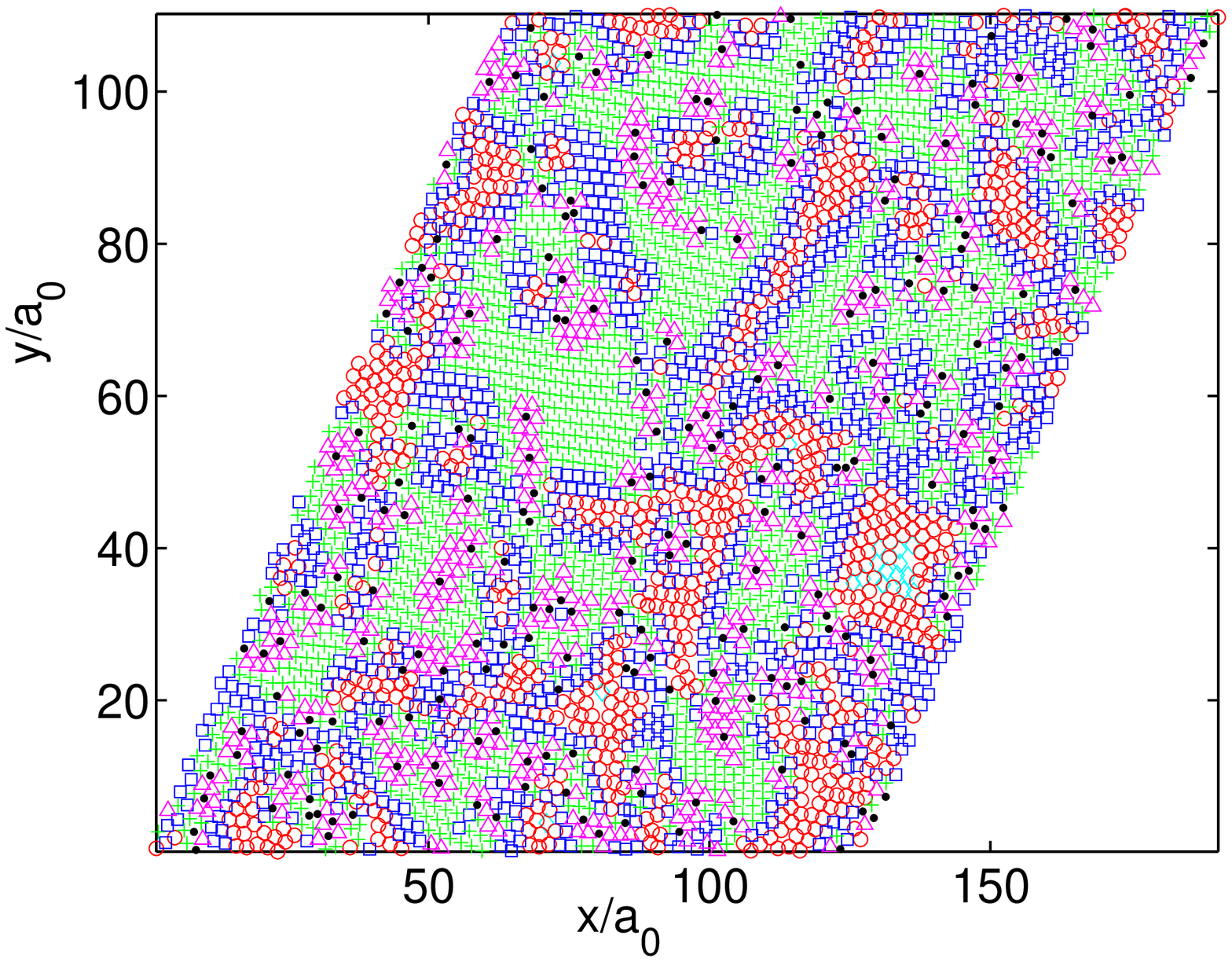}
\includegraphics[scale=0.4] {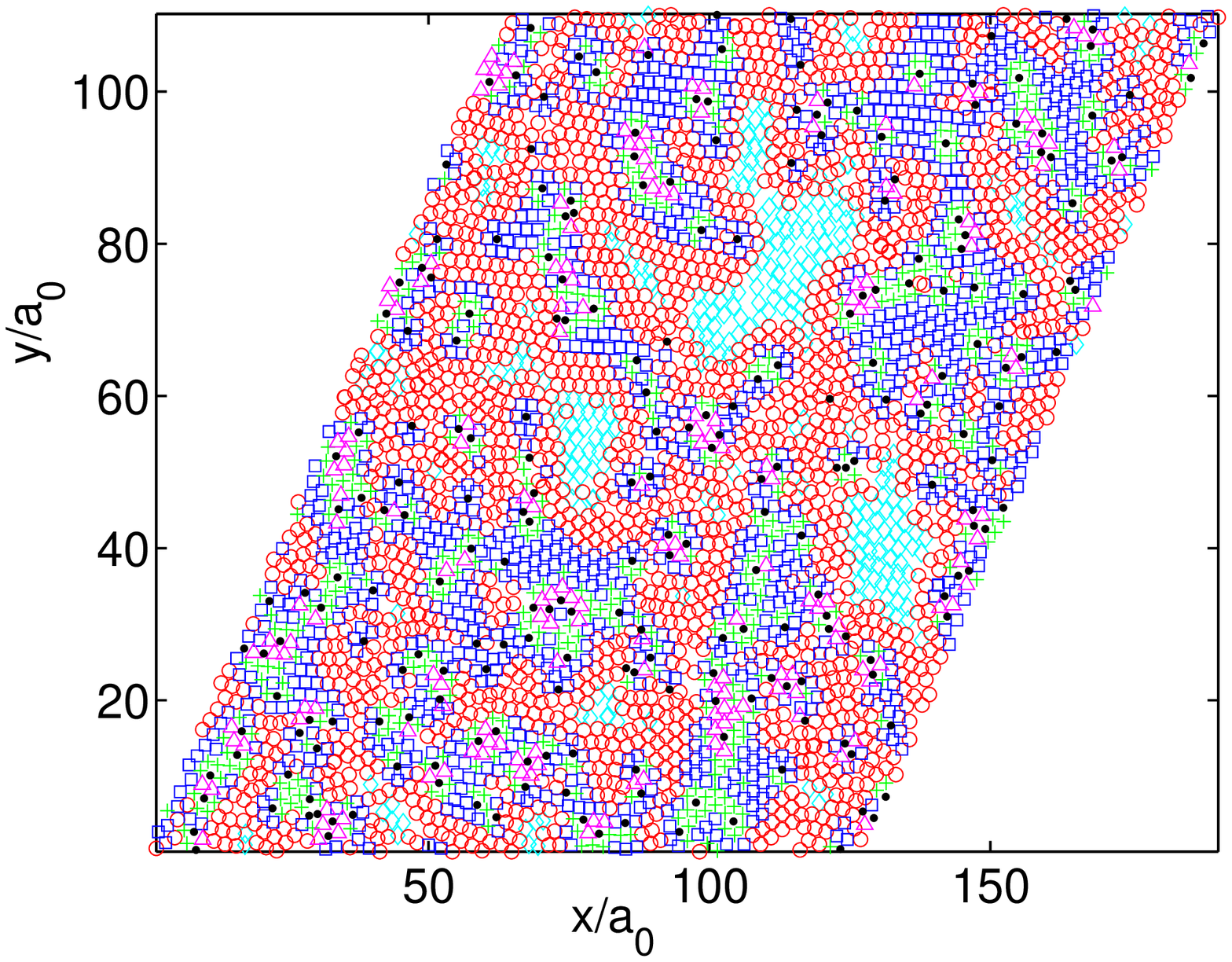}
\includegraphics[scale=0.4] {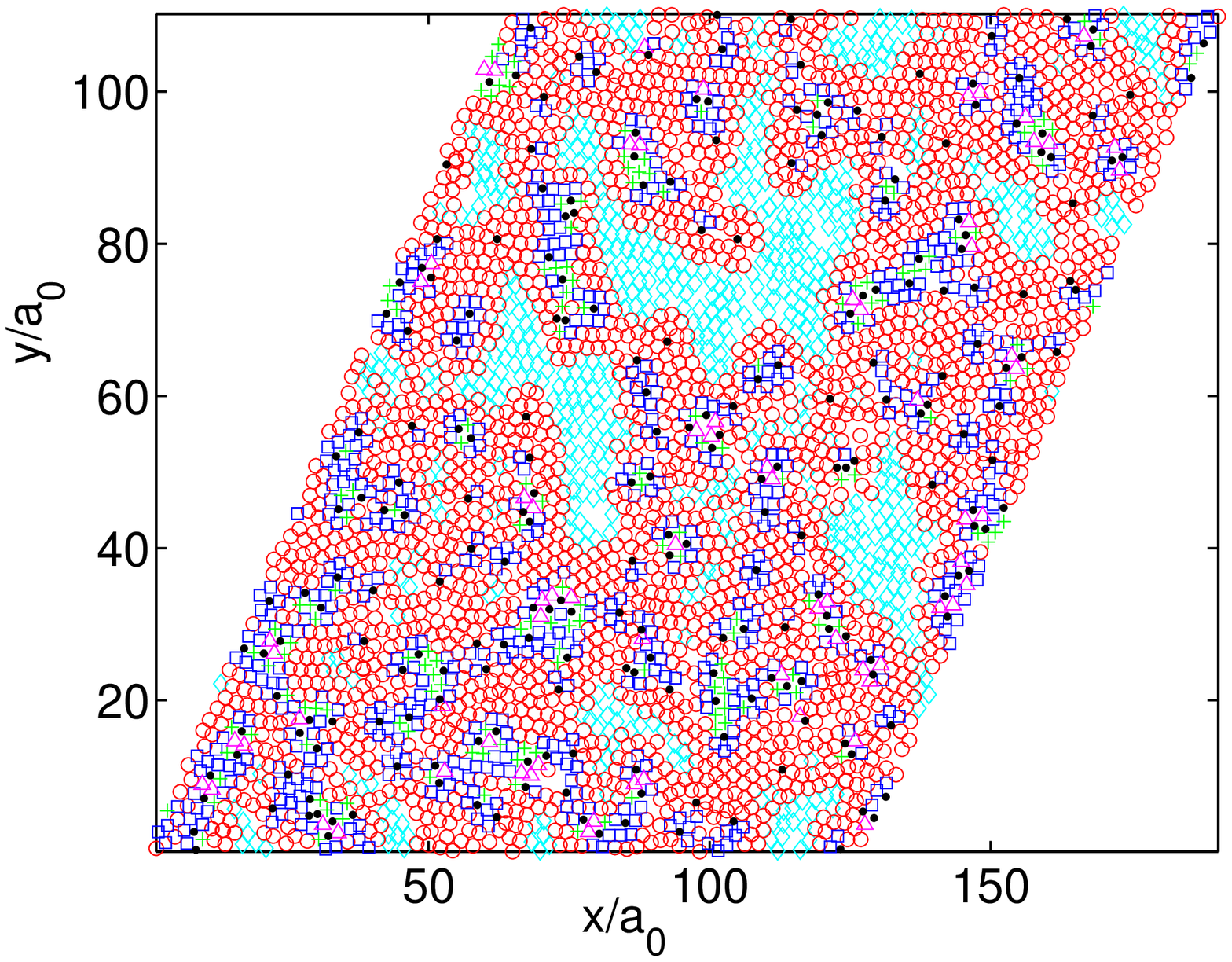}
\caption{\label{fig8} (Color online). Peak density plots for a sample
with $c=1/16$ (the same sample as in the top left panel of Fig.~\ref{fig7})
at temperatures (from top left to bottom) $T=17.6$K, $T=18.4$K and $T=19.2$K.
The locations of the peaks (see text)
are  displayed as data points, while the values of the density at these peaks 
are shown according to
the (color and) symbol coding: (cyan) diamonds:
peaks with $\rho_{\hbox{peak}}/\rho_0
< 1.5$, (red) empty circles: $1.5 \le \rho_{\hbox{peak}}/\rho_0 < 3.5$, (blue)
squares: $3.5 \le \rho_{\hbox{peak}}/\rho_0 < 5.5$,
(green) plus signs: $5.5 \le \rho_{\hbox{peak}}/\rho_0 < 7.5$, and
(magenta) triangles: $7.5 \le \rho_{\hbox{peak}}/\rho_0 < 9.5$. The
(black) asterisks denote the positions of pinning centers.}
\end{figure}

The same $c=1/16$ configuration 
studied in the top left panel
of Fig.~\ref{fig7} is examined in Fig.~\ref{fig8} as a function of 
the temperature. In this case we show plots of the vortex density at the
local density peaks, rather
than Voronoi plots, at three values of $T$, $T=17.6$K, $18.4$K and $19.2$K. 
The values $\rho_{\hbox{peak}}$ 
of the density at the local density peaks, normalized
to $\rho_0$, are shown according
to the symbol (and color) scheme explained in the caption. Higher values of
the peak density indicate strongly localized, solid-like behavior, while lower
values indicate that the 
corresponding vortices are more delocalized, as in a liquid state.
At the lowest temperature displayed, one can see rather large regions
where the vortices are indeed strongly localized. These regions are separated by
smaller regions  where the vortices are more delocalized. 
By comparing this figure (in particular the top right
panel) with the top left panel of Fig.~\ref{fig7}, one sees that the delocalized
regions are associated with high defect density, indicating that the vortices
lying near the grain boundaries are more delocalized than the ones lying
inside the crystalline grains. 
It is apparent that, as $T$ increases, the regions with higher values
of $\rho_{\hbox{peak}}/\rho_0$ shrink, as  the localized, more solid-like,
regions ``melt'' into more liquid-like areas. One can visualize, therefore,
that the ongoing transition process proceeds very locally, with different
regions in the sample behaving in different ways. This explains
the very rounded ``transition'' observed in the order parameter and free energy
plots (Figs.~\ref{fig6} and \ref{fig5} respectively).

\begin{figure}
\includegraphics[scale=0.4] {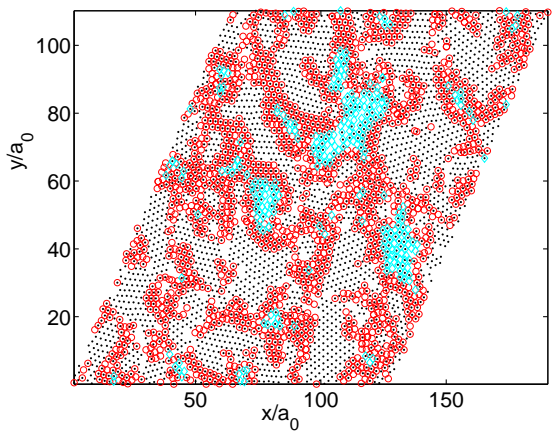}
\caption{\label{fig8p} (Color online) Relation between topological
defects and the value of the density peaks. The black dots mark the locations
of sixfold coordinated density peaks. The (cyan) diamonds indicate locations
where the peak density is low (less that $1.5\rho_0$) while the open (red)
circles represent locations with peak density between $1.5\rho_0$ and
$3.5\rho_0$. The temperature is $18.4$K, $c=1/16$ and the sample is the
same as in Fig.~\ref{fig7} and Fig.~\ref{fig8}.}
\end{figure} 

There is another way to observe the correlation between the defect structure
and the values of the local peak density in the vortex lattice, and this view
is shown in Fig.~\ref{fig8p}. There we consider again $c=1/16$ at $T=18.4$K.
This plot combines features of Figs.~\ref{fig7} and \ref{fig8}. The
same sample as in Fig.~\ref{fig8} is considered.
The six-fold
coordinated peak locations only (that is, the ones not associated with
defects) are shown as black dots, while  only the peak positions with 
low values of the peak density
are indicated by the symbol (and color) scheme. One can then clearly
notice that the local density peaks with low values of the peak 
density are mostly associated with peak positions that
are {\it not} six-fold coordinated. This plot clearly illustrates the 
strong correlation between the degree of localization and the presence
of topological defects: vortices lying near grain boundaries where the 
topological defects are concentrated are more delocalized than the ones
lying inside the crystalline grains. A similar correlation between the
degree of localization and the presence of defects has been observed in
several experimental and numerical studies of particle systems~\cite{colloid}.

\begin{figure}
\includegraphics[scale=0.35]{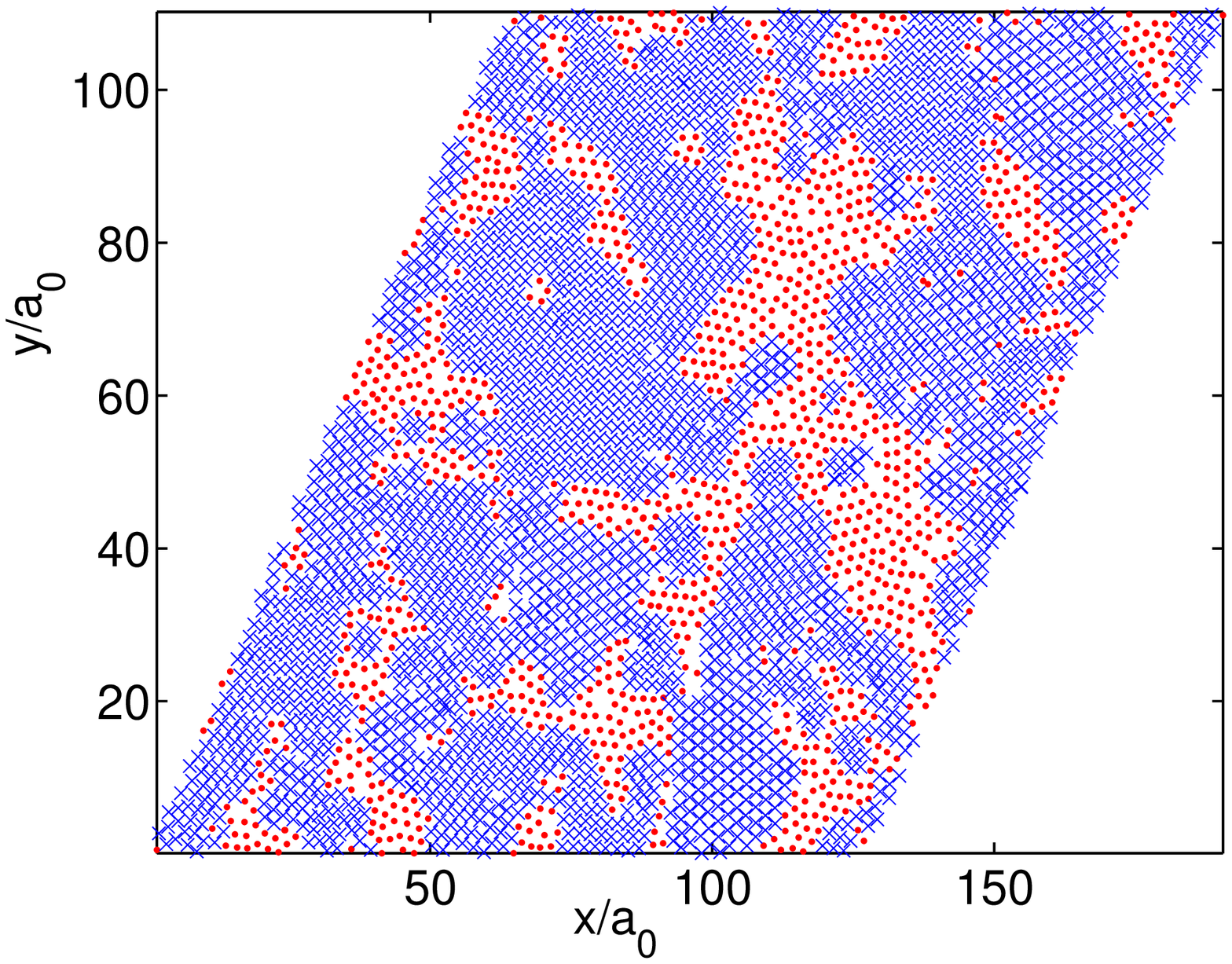}
\includegraphics[scale=0.35]{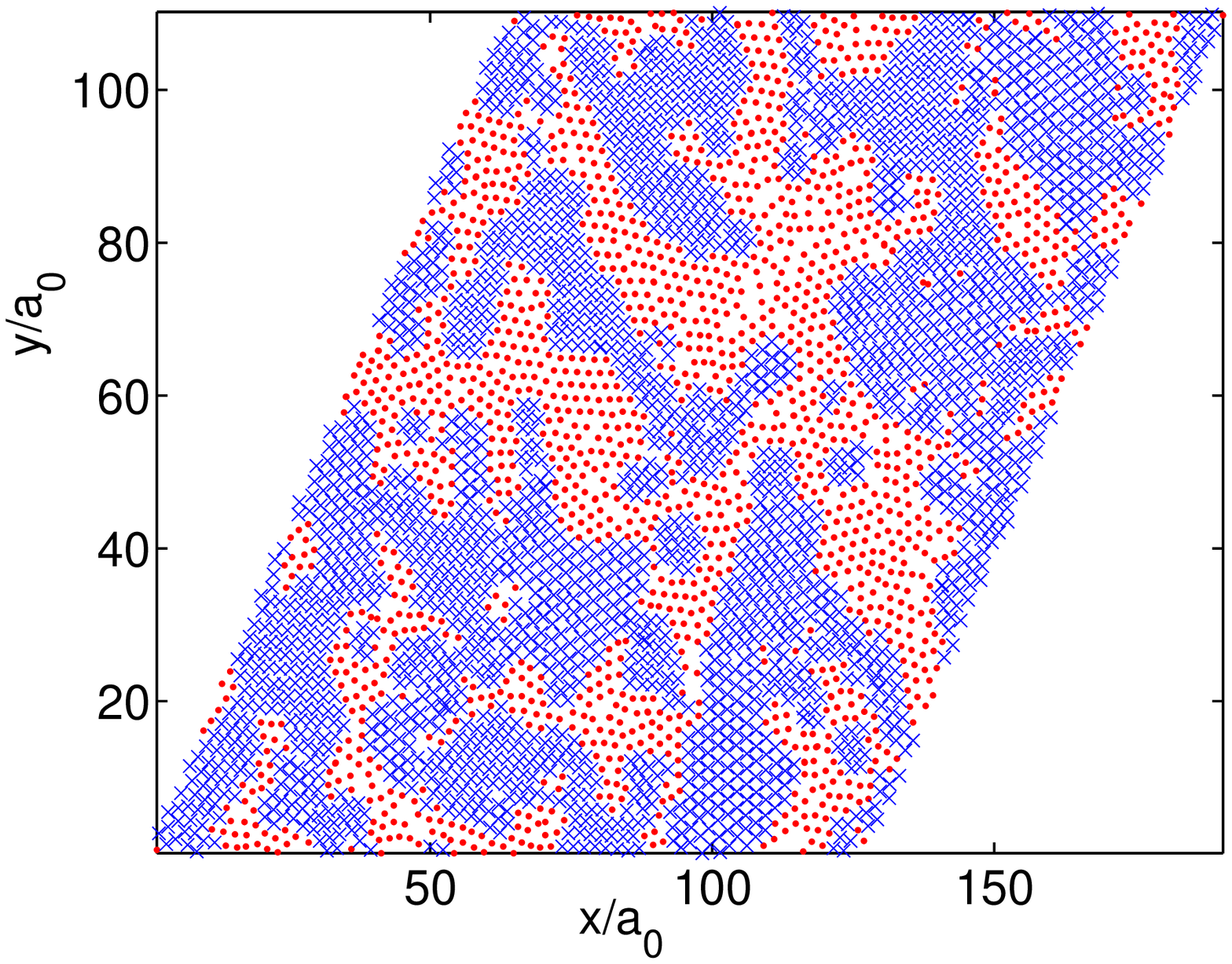}
\includegraphics[scale=0.35]{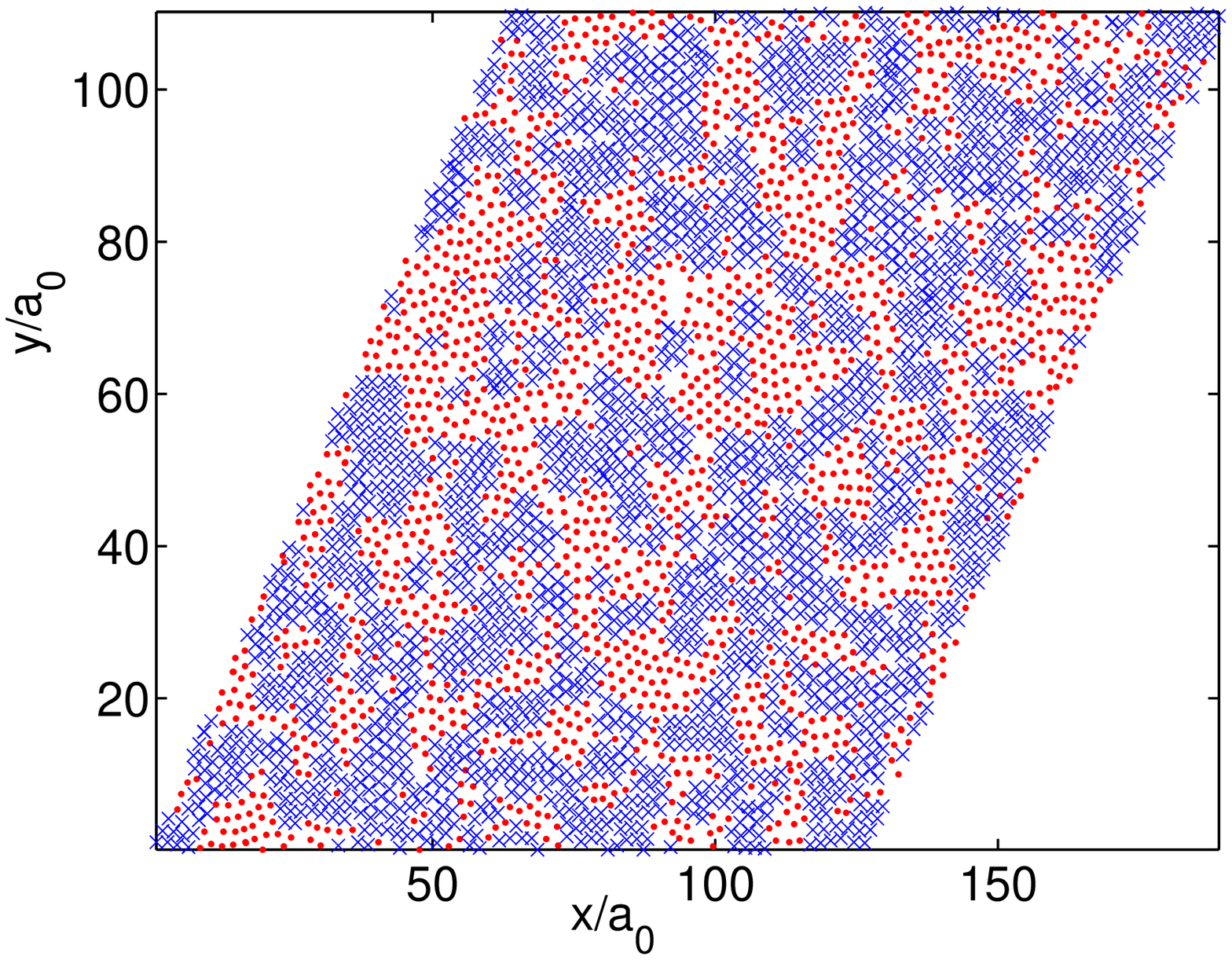}
\includegraphics[scale=0.35]{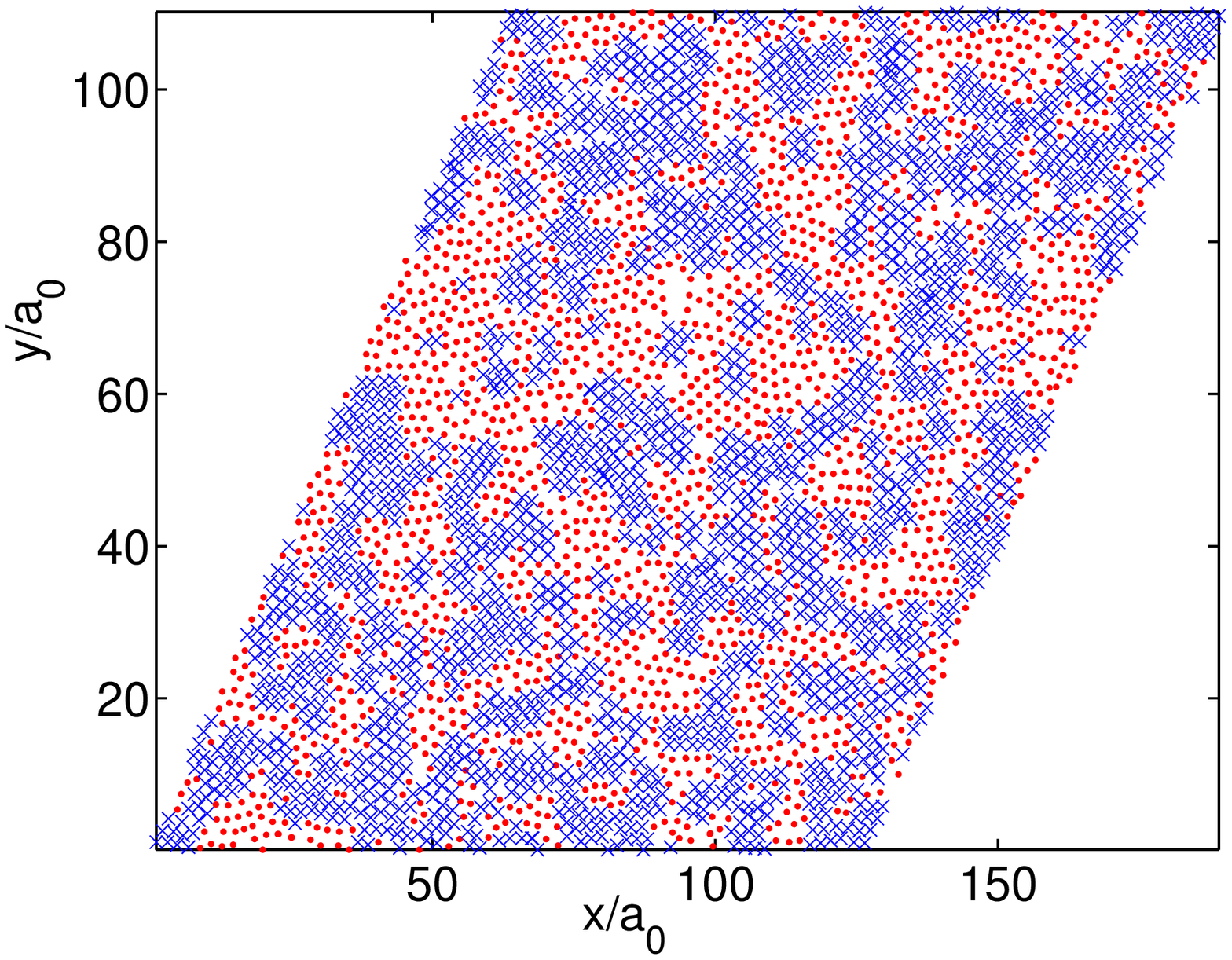}
\caption{\label{fig9} (Color online) Percolation plots. The (blue) crosses
represent regions where the system is solid-like (see text) while the (red)
dots indicate liquid-like areas. The two top panels are for $c=1/16$ at
$T=18.2$K (left side, where the solid percolates) and $T=18.4$K (right
side, where the liquid percolates). The two bottom panels are
for $c=1/8$ and $T=19.8$K (left) and $T=20.0$K (right). }
\end{figure}

\begin{figure}
\includegraphics[scale=0.5]{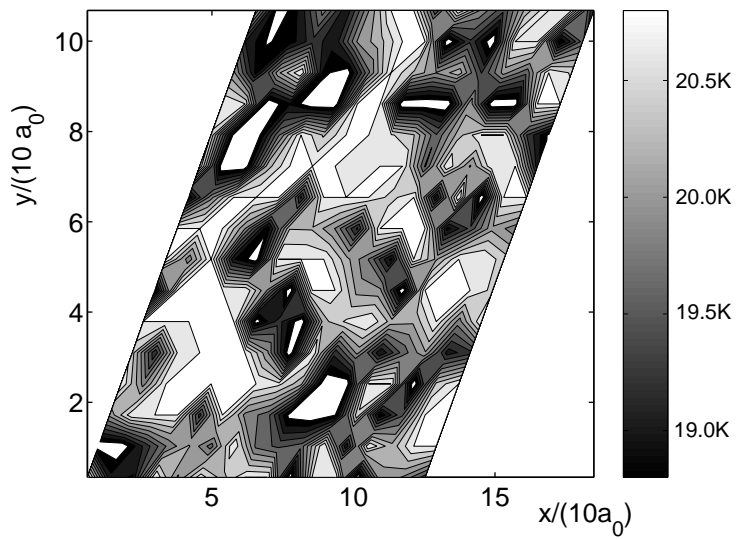}
\caption{\label{fig9p} Gray scale (as indicated) contour plot of the
local melting temperature as defined in the text. Results shown are for one sample 
with $c=1/8$ where percolation occurs near $T=20$K.}
\end{figure}

The question still remains, of how to characterize the 
smooth ``transition''
between BoG and IL phases for $c \geq 1/16$. Since the minimum representing 
the IL at high temperatures transforms continuously into the one 
corresponding to the BoG phase at low temperatures, we can not use 
a crossing of free energies to locate the transition temperature, as was
done for smaller values of $c$. 
We are here guided by two observations: one is that the transition 
is so broad because
it takes place locally at 
different temperatures in different regions of the sample. 
Second, it was noted in Ref.~\onlinecite{prbr} that at lower concentrations,
the BoG to IL melting transition coincides with a percolation process:
at lower $T$ the solid-like regions percolate throughout the sample, while
at higher $T$ it is the liquid-like regions that do. It was also noted there 
that one can define a local transition temperature using a
criterion based on the degree of localization of the vortices, 
that this local transition temperature
varies across each sample, and that the temperature at which regions
that have already melted percolate across the sample coincides with the 
global transition temperature 
determined from the crossing of the 
free energies of the minima representing the two phases. A similar connection
between the BoG--IL transition and percolation is suggested in a numerical
study~\cite{Shes95} of a disordered two-dimensional boson model to which 
the thermodynamic behavior of the system under study here can be mapped
approximately. 

We adopt here then this point of view and search for a percolation transition.
At concentrations larger than $c=1/32$, the temperature at which 
this transition occurs is then plotted as
the dashed melting curve of Fig.~\ref{fig1}.
Two examples of our search for percolation are shown in Fig.~\ref{fig9}. 
Two samples are considered,
one at $c=1/16$ the other at $c=1/8$. In the plots we identify the
local density peak sites where $\rho_{\hbox{peak}}/\rho_0 \le 3$  as being 
in a liquid  region, and those for which $\rho_{\hbox{peak}}/\rho_0 > 3$
as belonging to a solid-like region. This criterion was found~\cite{prbr}
to be appropriate in our studies of systems with  smaller values of $c$. 
We can see in Fig.~\ref{fig9} that a
percolation transition can indeed be located, and that the transition
temperature increases with $c$. The results are remarkably consistent across
samples of the same concentration. The results used to plot
this portion of the phase boundary (dashed line) in Fig.~\ref{fig1}
are averages over three to five samples.

A local transition temperature  can also be defined and the results correlated
with the percolation plot. To do this we divide the vortex lattice into small
regions containing about 16 vortices each. Within these regions we can 
calculate 
the average local peak density, $\rho_{av}^p$. In calculating this average,
we exclude the very high peaks associated with any existing pinning centers. 
The ``local melting temperature'' of each of the small
regions is then defined as the temperature at which $\rho_{av}^p$ for
that particular region drops (as the overall $T$ increases) from above
the threshold value of 
$3\rho_0$ discussed above, to below. An example is shown in Fig.~\ref{fig9p},
for a concentration $c=1/8$ in the upper limit of the range studied. This
is a gray scale plot in which the values of this local melting temperature
are shown according to the code indicated by the adjacent bar. Similar
spatial variations of
a local transition temperature have been observed in several experiments
\cite{Banerjee03,Soibel01}.
For this
sample, the overall transition, as determined by the percolation method,
occurs at $20$K. It is clear from the plot that the regions of the sample that
have melted at $T=20$K form a barely percolating cluster, indicating that
the result for the transition temperature obtained from these considerations is
consistent with that of the overall percolation analysis. 

\begin{figure}
\includegraphics[scale=0.4] {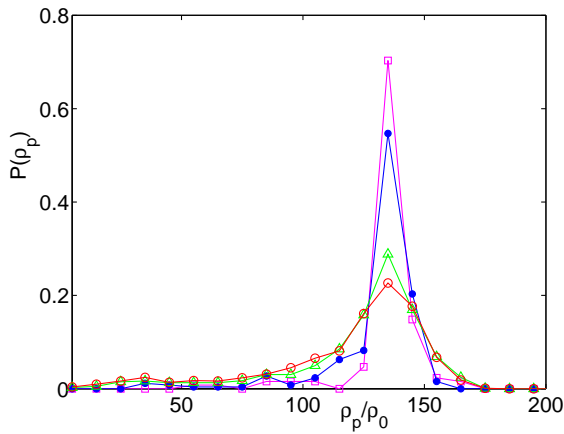} 
\caption{\label{fig10} (Color online). Distribution  of the peak density
$\rho_p$ (see text) evaluated at each computational
cell containing a pinning center for different values of $c$. The vertical
axis is the probability of having a certain value
of the normalized $\rho_p$. Results are shown
at $T=21.6$K for four different
concentrations: $c=1/64$ [(magenta) squares], 
$c=1/32$ [(blue) filled circles], $c=3/32$ [(green) triangles], and $c=1/8$
[(red) open circles]. Straight solid lines join successive data points}
 
\end{figure}

\subsection{Depinning crossover}\label{depin}

The above completes the discussion of the main portion of the phase diagram.
It remains to explain how the depinning temperature (the dotted line on the
right side of Fig.~\ref{fig1}) was determined. To discuss this point
we begin by noting that at sufficiently low $T$,  every pinning site
should hold one vortex. Indeed, as explained above, the strength and range
of the pinning potential were chosen so that this would be the case. One
can verify that this is so by adding, for a given
pinning center,  the values
of the  variables $\rho_i$ on the computational lattice, for all the sites $i$
which are within the pinning  range $r_0$ of  that pinning center.
This ``integrated density'' at a pinning center\cite{prbv}
is nothing but the time averaged 
number of vortices pinned at that center.  At low $T$ it tends 
to unity at nearly all centers, the rare exceptions occurring in some 
configurations 
at high $c$ where two sites may be very close to each other. 

For a given configuration, we can similarly compute for each pinning center
a quantity $\rho_p$, defined as the vortex density in
a computational cell that contains
a pinning center ($\rho_p \equiv \rho_i/(\sqrt{3} h^2/2)$ 
where $\rho_i$ is the discrete density
variable at computational cell $i$ which contains a pinning center, and $h$
is the spacing of the computational mesh). 
This quantity should not to be confused with 
$\rho_{\hbox{peak}}$ which refers to the density at 
local density peaks other than the ones
at pinning centers. It is very instructive to consider
the distribution of $\rho_p$ values (normalized to
$\rho_0$) as a function of $c$ and $T$. Such distributions for several
different values of $c$ at a fixed $T=21.6$K, near the 
depinning temperature, are shown in Fig.~\ref{fig10}.
These distributions are sharply peaked at a value which
is somewhat less than that corresponding to having one vortex in the cell
(one vortex in a computational cell 
corresponds to a value of about 234.9 for the 
normalized quantity $\rho_p/\rho_0$), regardless of $c$. 
The distribution broadens markedly with increasing $c$ and a small
secondary peak on
the low side develops. At these values of $T$ and $c$, 
the percentage of pinning sites
that hold only a small fraction of a vortex is clearly not negligible.

This behavior can be understood from a consideration of the effects of
neighboring pins on the probability of a particular pinning center being
occupied by a vortex. Due to the strong 
short-distance repulsive interaction between two
vortices, 
two pinning centers that lie 
very close to
each other can not be simultaneously occupied. As shown in 
Ref.~\onlinecite{prbv}, this interaction effect becomes important when the
separation between two pinning centers becomes smaller than about half the
average intervortex spacing. Specifically, it was shown there that for
values of $B$, $T$ and the pinning parameters $\alpha$ and $r_0$ very similar
to those used in the present calculation, two pinning centers separated by
less than $a_0$ are not simultaneously occupied. We find that all the
pinning centers for which the value of $\rho_p/\rho_0$ 
is less than 50 have at least another pinning center within a distance $a_0$
from them. The probability of finding such closely spaced pairs of
pinning centers obviously increases with increasing $c$. This is the reason for
the broadening of the distribution of $\rho_p/\rho_0$ on the low
side as $c$ is increased.
The pinning efficiency of a pinning center can also be enhanced by the
presence of a neighboring pin if the distance between the two pins is close
to the average intervortex spacing~\cite{prbv}. This is the reason for the
occurrence of the tail of the distribution of $\rho_p/\rho_0$ 
on the high side. However, the 
overall effect is a reduction in the average value of $\rho_p/\rho_0$ as the
pin concentration is increased.

\begin{figure}
\includegraphics [scale=0.65] {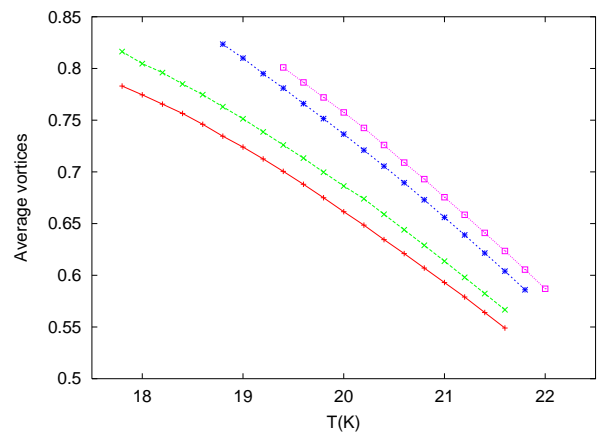}
\caption{\label{fig11}(Color online) The average number of vortices pinned 
at each pinning center, as a function of $T$, for four values of $c$:
(red) plus signs and solid line are for $c=1/8$, (green) crosses and long dashed
line, $c=3/32$, (blue) asterisks and short dashes, $c=1/32$, (magenta)
squares and dotted line, $c=1/64$.  }
\end{figure}

The integrated density at each pinning center can be averaged over all 
centers, and also over all pin configurations at a given value of $c$
and $T$ to obtain an overall average number of vortices pinned at each
pinning center at that concentration and temperature. This is of course
the quantity of main interest to determine the overall behavior
of the system. Results for this average number of vortices are plotted
in Fig.~\ref{fig11} as a function of temperature, for several
concentrations. The results are averages over four or five
samples, the dispersion being very small. 
At low $T$ the 
average number of vortices  tends towards unity, and 
it decreases as $T$ increases.
It is in general (see Fig.~\ref{fig11}) a function of $c$, decreasing as $c$
increases. This reflects the interaction effects discussed above and 
depicted in Fig.~\ref{fig10}.
In Ref.~\onlinecite{prbv} it was shown (see Fig.~3 of that work)
that the average number of vortices pinned by a single center, as a function
of the strength parameter $\alpha$ at constant $T$, 
increases sharply from a low
value to one close to unity as the value of $\alpha$ increases,
with  the middle of this sharp increase occurring at a 
value corresponding to the integrated density being close to 0.6. 
It is therefore appropriate
to associate the depinning temperature with the point where each curve
in Fig.~\ref{fig11} crosses the ``critical'' value of 0.6, and this criterion
has been used to determine the location of the 
depinning line in Fig.~\ref{fig1}. This criterion is very similar to the
one used in Ref.~\onlinecite{LV04} where the effects of other pinning centers
on the pinning properties of a particular one were not considered.
Consequently, Ref.~\onlinecite{LV04} predicts a depinning temperature that
is independent of $c$ for a fixed value of $B$.
The small reduction of the depinning 
temperature with increasing $c$ seen in Fig.~\ref{fig1}
reflects the interaction effects discussed
above, which reduce the pinning efficiency of individual pinning centers
as $c$ is increased. 

The depinning temperature has been experimentally measured~\cite{Banerjee04} 
in samples of BSCCO for several small concentrations of columnar pins. It is
found  that the depinning temperature for  fixed 
$B$ in the regime where the depinning
crossover occurs in the liquid state (the regime considered in our
calculations) 
is rather insensitive to the pin concentration $c$. This is in qualitative
agreement with our results. There is, however, a
small discrepancy: the experimental results~\cite{privcomm} 
show a small increase of the
depinning temperature with increasing $c$, whereas we find a small
decrease. This difference may be understood as arising from
the different criteria used to define
the depinning crossover, in the following way.

In the experiment~\cite{Banerjee04}, the depinning temperature 
is determined from the
difference between the transport currents 
in the pristine (without columnar pins)
and irradiated (with columnar pins) regions of
the sample. This difference arises because the pinning property of the
IL (which is found only in the irradiated parts of the sample) with both 
liquid- and solid-like regions is
different from that of the homogeneous vortex liquid that exists in the
pristine parts of the sample. Now, our study shows
that solid-like regions in the IL occur near clusters of 
columnar pins (see, e.g.
Fig.~3 of Ref.~\onlinecite{prbr}). The pinning property of such a solid-like
region depends both on the pinning efficiency of individual pins, and on
the size (number of pins) of the cluster of pins present in the region. The
average cluster size increases with increasing $c$. So, it is reasonable to
expect that the pinning efficiency of a typical solid-like region in the
IL would also increase with  $c$. Also, the fraction of solid-like
regions in the IL at a fixed temperature would be higher for larger values
of $c$. These considerations imply that the pinning properties of the
inhomogeneous IL would become similar to those of the homogeneous liquid
at a higher $T$ in samples with larger $c$. Thus,
this effect tends to increase the delocalization temperature obtained from
the current distribution as $c$ is increased. It may dominate over the small
decrease arising from the effect found in our calculation, thereby
producing a overall increase in the measured delocalization temperature as
the value of c is increased. A similar argument would also explain the
differences between our results for the depinning line and those
obtained in the simulations of Nonomura and Hu~\cite{NH04} who used the 
measured value of the tilt modulus to determine the depinning temperature. 
Since the depinning
is a crossover, not a thermodynamic phase
transition, it is not surprising that different criteria lead to slightly 
different results for the depinning temperature.

\section{Summary and discussion\label{con}}

We have reported here the results of a detailed investigation of
the phase diagram 
of a highly anisotropic layered superconductor with columnar
pinning in the temperature ($T$) -- pin concentration ($c$) plane. For small
values of $c$, we confirm the 
occurrence~\cite{prlr,prbr} of a topologically ordered 
BrG phase at low temperatures and a
two-step first-order melting of the BrG phase into an IL, with a small
region of intermediate polycrystalline BoG phase. In the IL, a fraction 
of vortices remains localized at the strong pinning centers. The IL
crosses over into a completely delocalized vortex liquid at a depinning
temperature that is higher than the temperature at which the BoG--IL
transition occurs. The BrG phase disappears
as the pin concentration is increased and the two-step melting found at low
pin concentrations is replaced
by a single continuous transition from a low-temperature BoG phase to a
high-temperature IL. This transition
corresponds to the onset of percolation of liquid-like regions
across the system. The temperature at which this 
percolation transition occurs increases
with increasing pin concentration, while the depinning temperature is
nearly independent of the pin concentration. Consequently,
the width of the temperature interval
in which the IL is found decreases
and eventually goes to zero when the pin concentration becomes
sufficiently high (slightly above 1/8).

Although the system we have considered 
here is three dimensional, since we do
take into account the electromagnetic interaction between vortices on different
layers, the correlated nature of the pinning produced by the columnar defects
makes it, in some sense, quasi two dimensional. 
The presence of a BrG phase in our phase diagram may 
then appear surprising in view
of the well-known result~\cite{zeng99} that a BrG phase does not exist in two
dimensions. 
It has been argued
\cite{gl95} that a BrG phase is unlikely to exist in a three dimensional
system with columnar pins. From our study of finite-size samples, we can not,
of course, rule out the possibility that 
for any non-zero value of the pin
concentration $c$, disorder-induced free dislocations would appear at 
length scales much larger than those accessible in our study. 
It is, however, very clear from our results (see results
above and also Refs.~\onlinecite{prlr,prbr}) 
that the structure
of the nearly crystalline minima we have found for small $c$ is qualitatively
different from that of the polycrystalline BoG minima found for the same values
of $c$. Thus, our conclusion about the occurrence of a first-order phase
transition between the polycrystalline BoG phase found near the melting 
transition and a much more ordered phase found at low temperatures 
for small $c$
would remain valid even if the nature of the nearly crystalline phase
turns out to be different from a true BrG.

Several noteworthy features found in our study have been observed in
experiments~\cite{Khaykovich98,Banerjee03,Menghini03,Banerjee04} on BSCCO with
small concentrations of random columnar pins:  Thus, our findings 
of the 
polycrystalline nature of the low-temperature BoG phase for small pin 
concentrations; the existence of the IL (called ``vortex nanoliquid'' in 
Ref.~\onlinecite{Banerjee04}) with a coexistence of pinned and delocalized 
vortices, which crosses over into a fully delocalized vortex liquid at a 
``depinning'' temperature; the existence
of a first-order transition between the IL 
and BoG phases for small values of
$c$, which becomes continuous as $c$ is increased keeping $B$ constant;
the increase of the BoG--IL transition temperature with increasing pin
concentration; and the spatial variations of a locally defined transition
temperature, all are in
agreement with experimental results. Our results are also similar to
those obtained in the simulation of Ref.~\onlinecite{NH04}, except for some 
small differences in the location of the depinning line which,
as discussed in section
\ref{depin}, may be attributed to the use of different criteria for determining 
the temperature at which the depinning crossover
occurs. 
The low-temperature BrG phase found here for small $c$ has not
 yet
been observed in experiments. As discussed in Ref.~\onlinecite{prbr}, this
may be due to the metastability of the BoG phase into which the vortex liquid
freezes as the temperature is decreased in the presence of a magnetic field.
Explorations of ways in which this metastability can be avoided would be
worthwhile. We note that a recent study~\cite{Moretti04} of grain
boundaries in vortex matter suggests that the melting of the vortex lattice
in unpinned systems can also be preceded by an intermediate polycrystalline
phase.

Our computations have been performed
for  a fixed value of the magnetic induction, $B$ = 2kG. 
The behavior of the system for other values of $B$ may be
inferred from our results by combining them with those of other existing
studies~\cite{menon,mahesh,sidhu}. As discussed in section~\ref{largec},
we find that for fixed $B$ the number of crystalline domains in a 
polycrystalline BoG minimum is proportional to the
pin concentration $c$.
This suggests that the typical size of the crystalline domains is determined
primarily by the arrangement of the pinning centers. If this is so then for
a fixed arrangement of the pinning centers, 
the typical number of vortices lying
inside a crystalline grain would be proportional to $B$. This
is precisely as observed in the
 above mentioned
experiments of Ref.~\onlinecite{Menghini03}. Earlier studies~\cite{menon} of 
the model of pancake vortices considered here indicate that interlayer vortex
correlations decrease as  $B$ is increased and the system 
becomes effectively two-dimensional (the melting 
temperature of the pure vortex lattice approaches that of the two-dimensional
system) in the limit of large $B$. Existing studies~\cite{mahesh,sidhu} of
the ground states (or low-lying local minima of the energy) of 
randomly pinned two-dimensional vortex arrays show 
polycrystalline structures similar to those found for 
the BoG phase in our work. We
therefore expect that the polycrystalline 
structure of the BoG phase would remain essentially unchanged
as $B$ is increased for a fixed concentration of pinning centers. Since the
melting temperature of the pure vortex lattice decreases~\cite{menon} with
increasing $B$, the temperature at which the BoG--IL transition occurs would be
expected to decrease as $B$ is increased at constant  pin concentration. 
Existing results~\cite{zeng99} about the absence of a true 
Bragg glass in two dimensions suggest that 
the BrG phase found for small $c$ in our study 
would not be present in the effectively two-dimensional 
large-$B$ limit. 

Numerical
studies~\cite{NH04,rodriguez} in which the vortex system is represented by a 
frustrated XY model do not find any clear evidence for the formation of 
polycrystalline structures, whereas all existing numerical 
studies~\cite{Sen01,mahesh,sidhu} in which the vortices are modeled as
interacting particles (or lines) indicate that the structure of the 
low-temperature disordered 
phase is polycrystalline, as found in
our study. The reason for this difference is not 
clear. It may be due to the smallness of the sample sizes ($\sim$ 100 vortex
lines in the simulation of Ref.~\onlinecite{NH04}) used in the studies based on
the frustrated XY model. A polycrystalline structure can be discerned clearly
if the crystalline grains are 
large enough (e.g. for $c$ = 1/64 and 1/32 in our
calculation). Such large grains would be observed in a simulation 
only if the sample size is
much larger than the typical grain size, which is probably not the case in the
simulations of Refs.~\onlinecite{NH04,rodriguez} for weak pinning. For stronger
pinning, the typical grain size would become smaller than the sample sizes
considered in these studies. However, in this limit, it would be difficult to
distinguish a polycrystalline structure from an amorphous one. 
This difficulty should be
clear from an inspection of a small part (containing $\sim$ 100 vortices) of 
the bottom panel of Fig.~\ref{fig7} (for $c=1/8$) for which the grain size is
small.

Quantitatively, this mean-field calculation
is expected to provide accurate results for the transition temperatures as long
as the transitions are strongly first-order. This is true for the transitions
that occur at small values
of $c$ in our phase diagram. Our predictions are 
quantitatively less reliable for larger 
values of
$c$ for which the transition is continuous. The method of locating
the transition temperature from the crossing of the free energies of 
different local minima of the free energy does not work when the transition is
not first-order. Strictly speaking, the alternative method of
identifying the transition point as the temperature at which liquid-like
regions percolate across the system cannot conclusively
determine whether there is 
a true continuous transition for large $c$, or just a crossover from a 
high-temperature liquid-like state to a low-temperature frozen state. We
assume that a continuous BoG--IL transition occurs in this regime
because such a transition is found in transport 
experiments and in numerical work in which the superconducting phase variables
are  included. These
aspects are not accessible in our work because the free energy we consider 
involves only the local
density of the vortices. However, we believe that our estimate of the 
BoG--IL transition temperature from the percolation criterion we have used is
reliable.  The reason
for this assertion is that a 
similar criterion works quantitatively well for the continuous transition
in disordered two-dimensional boson models to which the vortex problem we
consider here can be mapped. Also, transitions between liquid and glassy
phases in several other systems, such as colloidal systems~\cite{coll}
and thin polymeric films ~\cite{poly}, have been 
quite satisfactorily described in the past in terms
of percolation criteria similar to the one used here. It would be interesting
to test the validity of this description of the BoG--IL transition in
experiments in which magneto-optic techniques~\cite{Banerjee03,Banerjee04} 
can be used to map out the
local liquid- or solid-like character of different regions of the sample.

\section{Acknowledgments}

One of us (C.D.) wishes to thank S. Banerjee, X. Hu, 
T. Nattermann and E. Zeldov for
helpful discussions.


\begin{thebibliography}{99}
\bibitem{note} The free-energy minima we identify as representing the BrG
phase exhibit BrG-like properties over the length scales accessible in our
samples of finite size.
\end{thebibliography}
\end{document}